\documentclass[fleqn,usenatbib]{mnras}

\usepackage[T1]{fontenc}
\usepackage{ae,aecompl}

\usepackage{graphicx}	
\usepackage{amsmath}	
\usepackage{amssymb}	

\usepackage{subfig}

\usepackage{booktabs}

\usepackage{cleveref}
\crefname{section}{\S}{\S\S}
\Crefname{section}{\S}{\S\S}

\newcommand{\fstar}{$f_{*}$}

\newcommand{\Lya}{Ly-${\alpha}$}
\newcommand{\Lyb}{Ly-${\beta}$}
\newcommand{\Lyg}{Ly-${\gamma}$}

     \title[The subtlety of \Lya{} photons and the 21-cm signal]{The subtlety of \Lya{}  photons: changing the expected range of the 21-cm signal}

\author[Reis, Fialkov, and Barkana]{
Itamar Reis$^{1}$\thanks{E-mail: itamarreis@mail.tau.ac.il},
Anastasia Fialkov$^{2,3}$\thanks{E-mail: afialkov@ast.cam.ac.uk}
and Rennan Barkana$^{1}$
\\
$^{1}$School of Physics and Astronomy, Tel-Aviv University, Tel-Aviv, 69978, Israel\\
$^{2}$Institute of Astronomy, University of Cambridge, Madingley Road, Cambridge, CB3 0HA, UK\\
$^{3}$Kavli Institute for Cosmology, Madingley Road, Cambridge, CB3 0HA, UK\\
}

\date{Accepted XXX. Received YYY; in original form ZZZ}

\pubyear{2021}

\begin{document}
\label{firstpage}
\pagerange{\pageref{firstpage}--\pageref{lastpage}}
\maketitle

\begin{abstract}
    We present the evolution of the 21-cm signal from cosmic dawn and the epoch of reionization (EoR) in an upgraded model including three subtle effects of \Lya{} radiation: \Lya{} heating, CMB heating (mediated by \Lya{} photons), and multiple scattering of \Lya{} photons. Taking these effects into account we explore a wide range of astrophysical models and quantify the impact of these processes on the global 21-cm signal and its power spectrum at     observable scales and redshifts. We find that, in agreement with the literature,   \Lya{} and CMB heating raise the gas temperature by up to $\mathcal{O}(100)$ degrees in models with weak X-ray heating and, thus, suppress the predicted 21-cm signals. Varying the astrophysical parameters over broad ranges, we find that in the upgraded model the absorption trough of the global signal reaches a lowest floor of $-165$ mK at redshifts $z\approx 15-19$. This is in contrast with the predictions for a pure adiabatically cooling Universe, for which the deepest possible absorption is a monotonically decreasing function of cosmic time and is $-178$ mK  at $z = 19$ and $-216$ mK at $z=15$, dropping to even lower values at lower redshifts (e.g. $-264$ mK at $z = 10$). With the \Lya{} and CMB heating included we also observe a strong suppression in the low-redshift power spectra, with the maximum possible power (evaluated over the ensemble of models) attenuated by a factor of $6.6$ at $z=9$ and $k = 0.1$ Mpc$^{-1}$.  Finally, we  find that at high redshifts corresponding to cosmic dawn, the heating terms have a subdominant effect while  multiple scattering  of \Lya{}  photons is important,  leading to an amplification of the power spectrum by a factor of $\sim 2-5$.

\end{abstract}




\section{Introduction}
\label{sec:introduction}

Neutral hydrogen existing in the  intergalactic medium (IGM) prior to the completion of the process of reionization  produces radio signal at the intrinsic wavelength of 21 cm. The intensity of this signal depends
 on the radiative outputs of the first stars and galaxies as well as on the cosmological parameters \citep[e.g.,][]{madau97,barkana18book,Mesingerbook}. Owing to these dependencies, observations of the redshifted 21 cm signal with interferometers such as the  Square Kilometer Array \citep[SKA,][]{koopmans15}, the New Extension in Nan\c{c}ay Upgrading LOFAR  \citep[NenuFAR,][]{zarka12} and the Hydrogen Epoch of Reionization Array \citep[HERA,][]{deboer17}, and with radiometers such 
as the Experiment to Detect the Global EoR Signature \citep[EDGES,][]{bowman18}, the Shaped Antenna measurement of the background RAdio Spectrum \citep[SARAS,][]{Singh:2018},  the Large-Aperture Experiment to Detect the Dark Ages \citep[LEDA,][]{Price:2018},  the Mapper of the IGM Spin Temperature\footnote{http://www.physics.mcgill.ca/mist/} (MIST), and the Radio Experiment for the Analysis of Cosmic Hydrogen\footnote{https://www.kicc.cam.ac.uk/projects/reach} (REACH), are bound to provide an unmatched view of the structure of the Universe over the epochs referred to as cosmic dawn and reionization (cosmological redshifts of $z\sim 6-35$). 

Theoretical modeling of this signal is challenging as it depends on a large number of non-linear and non-local phenomena. A major role is played  by   the cumulative emission of ultraviolet (UV)  and X-ray sources which affect the 21-cm signal within their respective effective horizons of up to a few hundreds comoving Mpc   \citep[e.g.,][]{fialkov14}.  Even though current state-of-the-art simulations of the 21-cm signal   \citep[e.g,][]{santos10, mesinger11, visbal12, fialkov14a, reis20a, reis20c}  include major processes that affect this observable, such as the Wouthuysen-Field effect \citep[WF, ][]{wouthuysen52, field58}, heating of the IGM by X-ray sources and the process of reionization by UV sources,  not all of the more subtle effects are accounted for and their importance is not explored for the multitude of possible high-redshift astrophysical scenarios. In this work we focus on some of the more subtle effects of the \Lya{} radiative background.

\Lya{} photons play one of the most important roles in the 21-cm cosmology. The well-known WF effect, in which the absorption and re-emission of ambient \Lya{}  photons by the hydrogen atoms drives their spin temperature 
towards the kinetic gas temperature,   renders  the 21-cm line visible against the background radiation  \citep{madau97,barkana18book,Mesingerbook}. Stellar Lyman photons redshifted to the vicinity of the \Lya{} line experience multiple scattering \citep{semelin07, naoz08,  baek09, visbal18, molaro19, reis20a} which affects their spatial distribution, and, thus, affect the resulting 21-cm signal. The effect of multiple scattering  is rarely fully included in the 21-cm simulations. In particular,  in our past works we accounted approximately for the redistribution of the  photons due to their scattering in the wings of the \Lya{}  line \citep{naoz08} and assumed that photons travel in straight lines from emission until absorption \citep{FialkovLya}.  

In addition to the WF effect,  \Lya{} photons play yet another role by affecting the kinetic temperature of the gas through the same process of absorption and re-emission  \citep{furlanetto06b, chuzhoy07}. It was shown that {\it injected} \Lya{}  photons (i.e., photons that are cascading into the \Lya{}  line from higher Lyman levels) can cool the gas, while  {\it continuum} \Lya{}  photons (i.e., photons that are redshifted into the \Lya{} line)  heat the gas, with a net effect that depends on the gas temperature (more details are given below).

Cosmic Microwave Background (CMB) heating \citep{venumadhav18} is another heating mechanism which is often neglected. This mechanism involves energy transfer between the radio background photons and the hydrogen gas, mediated by \Lya{} photons. This heating term is important when X-ray heating is weak or absent, e.g. \citet{venumadhav18} reported a $\sim 10\%$ effect at $z\sim 17$ on the temperature of the gas in the adiabatic cooling case and with the CMB as the background radiation. Since this mechanism was only recently introduced, it is not included in most calculations of the 21-cm signal\footnote{However, it is included in some of our most recent works \citep{fialkov19, reis20c}.}. In the literature, the effect of \Lya{} and CMB heating on the thermal state of the gas was predominantly considered in scenarios without X-ray heating  \citep{chuzhoy07, venumadhav18, ghara19, mittal20} due to the assumption that, whenever present, X-ray sources dominate  the  heating budget.

In this work we explore the impact on the global 21-cm signal and fluctuations  of the subtle  \Lya{} effects that are often neglected, namely (i) the effect of multiple scattering of \Lya{} photons  on the spatial distribution of these photons and, thus, the WF coupling \citep{semelin07, naoz08,  baek09, visbal18, molaro19, reis20a}, (ii)  the  contributions to the IGM heating  of   \Lya{} photons \citep{madau97} and (iii) heating of the IGM by the CMB photons \citep{venumadhav18}. We  assess the importance of these  contributions over the vast allowed astrophysical landscape. 

This paper is organized as follows: In \cref{sec:simulation} we describe our semi-numerical code for calculating the 21-cm signal. In   \cref{sec:lya_calculation} we focus on the effects explored in this work: \cref{sec:lya_ms_calculation} details the calculation of the spatial distribution of \Lya{} photons including multiple scattering, and \cref{sec:lya_heating_intro} describes the calculation of the \Lya{} heating effect, while we follow \citet{venumadhav18} in the implementation of the CMB heating. In \cref{sec:results} we present the results, exploring the effects of multiple scattering, \Lya{} heating, and the CMB heating for several representative cases. In \cref{sec:ranges} we show the revised landscape of the 21-cm signals (both global and power spectra) and compare it to our previous works \citep[e.g.,][]{cohen17, cohen18, cohen19}. In particular, {\em Fig. \ref{fig:ps_env} shows the most important result of this work}, namely the updated range of possible 21-cm signals in standard astrophysical models, for both the global signal and the 21-cm power spectrum. We summarize our main conclusions in \cref{sec:summary}. Throughout this paper we adopt cosmological parameters from \citet{Planck:2014}. All scales and wavenumbers are in comoving units.

\section{Simulating the 21-cm signal}
\label{sec:simulation}

\subsection{Formalism}

The 21-cm brightness temperature is given by \citep{madau97, barkana18book, Mesingerbook}
\begin{equation}
    T_{\rm 21} = \frac{T_{\rm S} - T_{\rm rad}}{1+z} \left( 1- e^{-\tau_{21}} \right),
\end{equation}
where $T_{\rm S}$ is the spin temperature, $T_{\rm rad}$ is the temperature of the background radiation which is usually assumed to be the CMB with   $T_{\rm rad} = T_{\rm CMB} = 2.725(1+z)$ K \citep[see][for the case with excess radio background above the CMB]{feng18,ewall18, fialkov19, reis20c} and $\tau_{21}$ is the 21-cm optical depth
\begin{equation}
\label{eqn:tau21}
    \tau_{21} = \frac{3 h A_{10} c \lambda_{21}^2 n_{\rm H} }{32 \pi k_{\rm B} T_{\rm S} (1+z)  (dv_r/dr) },
\end{equation}
where $h$ is the Planck constant, $A_{10}$ is the spontaneous decay rate of the hyper-fine transition,  $c$ is the speed of light, $\lambda_{21}=21.1$ cm is the wavelength of the 21-cm line, $n_{\rm H}$ is the neutral hydrogen number density, $k_{\rm B}$ is the Boltzmann constant, $dv_r/dr = H(z)/(1 + z)$ is the gradient of the comoving velocity field along the line of  sight and  $H(z)$ is the  Hubble rate.

The spin temperature is calculated as follows
\begin{equation}
\label{eqn:ts}
    T_{\rm S} = \frac{x_{\rm rad} + x_{\rm tot}}{x_{\rm rad} T_{\rm rad}^{-1} + x_{\rm tot} T_{\rm K}^{-1}},
\end{equation}
where
\begin{equation}
\label{eqn:xrad}
    x_{\rm rad} = \frac{1-e^{-\tau_{21}}}{\tau_{21}},
\end{equation}
and $T_{\rm K}$ is the kinetic temperature of the gas. The coupling  coefficient $x_{\rm tot}$ is the sum of the contributions of \Lya{} photons ($x_{\alpha}$) and collisions ($x_c$): $x_{\rm tot} = x_{\alpha} + x_{\rm c}$, with the  \Lya{}  coupling coefficient \citep{barkana18book}
\begin{equation}
    x_{\alpha} = \frac{1}{A_{10} T_{\rm rad}} \frac{16 \pi^2 T_{\ast} e^2 f_{\alpha}}{27 m_{e} c} J_{\alpha},
\end{equation}
where $f_{\alpha} = 0.4162$ is the oscillator strength of the \Lya{} transition, $T_{*} = 0.0682$ K and $J_{\alpha}$ is the local intensity of the \Lya{} radiation field. The  coupling coefficient due to collisions is
\begin{equation}
\label{eqn:col}
    x_c = \frac{1}{A_{10} T_{\rm rad}} \kappa_{1-0}\left(T_{\rm K}\right) n_{\rm H} T_{\ast}, 
\end{equation}
where   $\kappa_{1-0}\left(T_{\rm K}\right)$ is a known atomic coefficient \citep{allison69, zygelman05}. Note that $x_c$ is expected to be small at the redshift considered in this work \citep{barkana18book}. 

Eqns. \ref{eqn:tau21}, \ref{eqn:ts}, and \ref{eqn:xrad} present a recursive relation satisfied by $T_{\rm S}$ and $x_{\rm rad}$. In our simulation we do not rely on  the smallness of the optical depth and solve these Eqns. iteratively until convergence of $x_{\rm rad}$ and $T_{\rm S}$, starting from $x_{\rm rad}$ = 1 \citep{fialkov19}. Deviation of $x_{\rm rad}$  from unity is caused by stellar \Lya{} photons which transfer energy between the radio background and the thermal motions of the gas atoms, thus heating up the gas. This, in essence, is the  CMB  heating mechanism  recently pointed out by
 \citet{venumadhav18} and  included in our simulation.

\subsection{Cosmic history} 

The 21-cm signal of neutral hydrogen is predicted to be a powerful  probe of cosmic history between the moment of thermal decoupling of the gas from the CMB at redshift  $z_t \sim 136$
\citep{barkana18book} and the end of the process of reionization which leads to the disappearance of the 21-cm signal at $z\sim 6$ \citep[e.g.][]{Weinberger:2019}.

After thermal decoupling, the gas temperature drops below  the CMB temperature  due to the faster adiabatic cooling of the gas compared to that  of  the radiation. Due to the high density of the gas and in the absence of stars (and, thus, \Lya{} photons), the spin temperature is coupled to the  gas temperature  through inter-atomic collisions, a process  which results in an absorption trough in the spectrum of the redshifted 21-cm signal.  As the universe expands,  the effectiveness of the collisions-induced coupling ($x_c$) decreases. If not for the \Lya{} coupling, the 21-cm signal would vanish at $z\lesssim 30$ owing to the proximity of the spin temperature to that of the background radiation.  

The \Lya{} photons produced by the first stars enable the WF coupling of the  spin temperature  to the gas  temperature. As still $T_{\rm K} < T_{\rm rad}$, this coupling imprints  another absorption trough in the spectrum of the redshifted 21-cm signal. The  \Lya{} coupling saturates when  $x_{\rm tot} \gg x_{\rm rad}$ leading to $T_{\rm S} \sim T_{\rm K}$.

Cosmic heating is another process that is enabled by star formation. During the course of cosmic history the neutral hydrogen gas is heated via several channels,  causing  the absorption signal to fade and, in some cases, leading to an emission feature in the spectrum of the 21-cm signal at high frequencies. A major source of cosmic heating is thought to be the high-redshift population of X-ray binaries \citep[XRBs,][]{fragos13}. Indeed, X-ray photons emitted by XRBs at energies below $\lesssim 1$ keV can successfully heat the hydrogen gas if produced in large quantities; however,   X-ray efficiency of high-redshift XRBs is unknown. If heating by high-redshift XRBs were inefficient, other heating mechanisms, such as  \Lya{} and  CMB heating, could be important.

Evolution of the hydrogen kinetic temperature, including heating by \Lya{}  (which we describe in detail in Section \ref{sec:lya_heating_intro}) and by the CMB photons, is given by
\begin{multline}
\label{eqn:heating_rate}
    \frac{dT_{\rm K}}{dz} = 2 \frac{T_{\rm K}}{1+z} + \frac{2 T_{\rm K}}{3 (1+ \delta_b)}\frac{d \delta_b}{dz} - \frac{d x_{\rm e}}{dz} \frac{T_{\rm K}}{1+x_{\rm e}} -\\ \frac{2}{3 k_{\rm B} (1 + f_{\rm He} + x_{\rm e} )} (\epsilon_{\rm X} + \epsilon_{\rm Compton} +  \epsilon_{\rm Ly_{\alpha}}+\epsilon_{\rm rad}),
\end{multline}
where $x_{\rm e}$ is the free electron fraction, $f_{\rm He}$ is the helium fraction, and $\delta_b$ is the baryon over-density.
Heating and cooling rates in the right hand side of the above equation are contributed by Hubble expansion, structure formation, change in the number of ionized particles, X-ray photons \citep[see][for $\epsilon_{\rm X}$]{mesinger11}, Compton \citep[see][for $\epsilon_{\rm Compton}$]{mesinger11},   \Lya{} photons (see Eqn. \ref{eqn:lya_heating_rate} for $\epsilon_{\rm Ly_{\alpha}}$) and the CMB, $\epsilon_{\rm rad}$, respectively.  The CMB heating  rate was derived by \citet{venumadhav18} and is given by
\begin{equation}
\label{eq:epsrad}
    \epsilon_{\rm rad} = \frac{x_{\rm HI} A_{10}}{2 H(z)} x_{\rm rad} \left(\frac{T_{\rm rad}}{T_{\rm S}} - 1\right)\frac{T_{\ast}}{T_{\rm K}}.
\end{equation}
$\epsilon_{\rm rad}$ is non zero when $T_{\rm S} \neq T_{\rm rad}$, that is, when some coupling between the spin temperature and the kinetic temperature has been induced. As we show below \citep[and as was first pointed out by ][]{venumadhav18}  this mechanism alone  has a $\sim 10\%$ effect on the temperature of the gas  (at $z\sim 17$) in the absence of X-ray and  \Lya{}  heating and assuming the CMB as the background radiation.

Contrary to reionization which is constrained by observations of high redshift quasars, galaxies and the CMB \citep[e.g.][]{Weinberger:2019}, the timing of \Lya{} coupling and cosmic heating are  unknown due to the lack of direct observations and large uncertainties in the properties of primordial radiative sources. Owing to the relatively low number of \Lya{} photons required for efficient \Lya{} coupling \citep{Mesingerbook}, this event is thought to follow shortly after the onset of star formation and well before the processes of reionization and heating occurred on large cosmological scales. Perhaps the most enigmatic process of all is heating of the IGM as the large uncertainty in the nature of heating sources and their properties remains.

\subsection{Modelling}

We use our own semi-numerical approach to calculate both the global signal and the large scale power spectrum of spacial fluctuations in the 21-cm signal \citep[e.g.][]{visbal12, fialkov14a, cohen16}. Each simulation is initialized with a set of initial conditions (IC) for density and relative velocity between dark matter and gas \citep[v$_{\rm bc}$,][]{tseliakhovich10}. 
In order to obtain a population of galaxies, we apply a modified Press-Schechter model \citep{press74, sheth99, barkana04} where the local abundance of dark matter halos is biased by both the  local large-scale density and v$_{\rm bc}$ \citep{fialkov12}. We assume that stars  form in dark matter halos with circular velocity above a threshold value, $V_c$. Star formation  efficiency, \fstar{},  is constant for galaxies with circular velocity above the atomic cooling threshold ($V_c = 16.5$ km s$^{-1}$), while for lower values of $V_c$ (but above the molecular cooling threshold of $4.2$ km s$^{-1}$) we include a logarithmic suppression of $f_*$ as a function of halo mass  \citep[e.g.][]{cohen19}. Finally, 
we take into account the effect of v$_{\rm bc}$ \citep{fialkov12}, Lyman-Werner feedback \citep{haiman97,fialkov2013}, and photoheating feedback \citep{rees86, sobacchi13,cohen16} on the minimum mass of star forming halos. In addition, in some parts of this work (Section \ref{sec:ranges}) we include Poisson fluctuations recently added to the code \citep{reis20a}.

The simulated population of galaxies produces X-ray, UV,  \Lya{} and Radio radiative backgrounds that drive the 21-cm signal.  For simplicity, we switch off the latter contribution in this work as the main focus of this paper is on the \Lya{} physics and the interplay between  \Lya{}, CMB and X-ray heating. The \Lya{} field emitted by stars is described in detail in \cref{sec:lya_ms_calculation}.
In the calculation we assume population II stars and use stellar spectra from \citet{barkana05}. We also take into account the contribution of \Lya{} photons produced via X-ray excitation of hydrogen atoms \citep[e.g.,][]{mesinger11} to the total  \Lya{} background when calculating the WF coupling coefficient.

The X-ray luminosity of galaxies is assumed to scale with the star formation rate \citep[SFR, e.g.,][]{Mineo:2012, fragos13, fialkov14a, Pacucci:2014}. We assume a constant efficiency factor $f_{X}$, which is a free parameter of the simulation (with $f_X = 1$ corresponding to present day galaxies). In addition to luminosity, spectral energy distribution (SED) of X-ray photons was shown to have an important effect on the 21-cm signal \citep{fialkov14a}. In our simulation we consider two possibilities: a power law SED parametrized by the slope, $\alpha$, and the low-energy cutoff, $E_{\rm min}$, and a realistic SED  derived by \citet{fragos13} for a population of high-redshift XRBs. 

The IGM is ionized by the UV (ionizing) photons with mean free path, R$_{\rm mfp}$ \citep[e.g.,][]{greig15}, produced by galaxies with ionizing efficiency $\zeta$  \citep[we use excursion set formalism,][]{furlanetto04}.  As was outlined in detail by \citet{cohen19}, given the values of other simulation parameters, there is one-to-one correspondence between $\zeta$ and the total CMB optical depth, $\tau$. Since only the latter is constrained by observations \citep[e.g.,][]{planckcollaboration18}, we use $\tau$ rather than $\zeta$ in the astrophysical parameter scans.

In total, here we employ a seven-parameter model with the parameters listed in Table \ref{tab:parameters} along with a brief descriptions and the ranges across which each parameter is allowed to vary. 

\begin{table}
\begin{center}

\begin{tabular}{l|ccc}

\toprule
Parameter & Allowed range  & Description     \\
\midrule
$f_*$ & $0.001 - 0.5$  &  Star formation efficiency \\
$V_c$ & $4.2 - 100$ km s$^{-1}$  & Minimum circular velocity\\
$f_X$ &  $10^{-4} - 10^{2}$  & X-ray production efficiency\\
$\alpha$ &  $1 - 1.5$  & X-ray SED slope\\
$E_{\rm min}$ &  $0.1 - 3$ keV  & X-ray SED minimum energy\\
$\tau$ &  $0.033 - 0.089$  & Optical depth to reionization\\
R$_{\rm MFP}$ &  $10 - 70$ Mpc  & Mean free path of ionizing photons\\

 \bottomrule
\end{tabular}
\caption{Astrophysical parameters  and their allowed ranges. Note that in models used in some of the figures we assume X-ray sources with a realistic predicted SED of X-ray binaries \citep{fragos13}, and those models do not include the parameters $\alpha$ and $E_{\rm min}$.} 
\label{tab:parameters}

\large
\end{center}
\end{table}

\section{The \Lya{} intensity field}
\label{sec:lya_calculation}

We now focus on the stellar contribution to the local \Lya{} background which enables the WF effect\footnote{When calculating the WF coupling coefficient we also add the contribution of \Lya{} photons generated by X-ray excitation of hydrogen atoms to the total  \Lya{} background. }. 
These photons are produced by  surrounding  galaxies and are originally emitted at frequencies between \Lya{} and Lyman limit.
There are two main channels to obtain local \Lya{} photons: (i) Photons produced at frequencies between  \Lya{} and  \Lyb{}  are   redshifted directly to  the \Lya{} frequency by cosmic expansion (continuum photons),  and (ii) photons emitted at higher frequencies are absorbed in higher  Lyman series frequencies  and create atomic cascades (injected photons).  About $30 \%$ of  cascades originating from the \Lyg{} and higher transitions produce  \Lya{} photons, while no  \Lya{} photons are produced by cascades originating from \Lyb{} \citep{hirata06, pritchard06}. We denote the  chance of producing a \Lya{} photon in a cascade from level $n$ by $f_{\mathrm{recycle}}(n)$ and assume that the cascade is instantaneous.

Consider photons absorbed at $z_{\rm abs}$ in frequency $\nu_n$. Such photons were emitted at redshift  $z_{\rm em}$ with frequency
\begin{equation}
    \nu'_n(z_{\rm abs}, z_{\rm em}) = \nu_n \frac{a_{\rm abs}}{a_{\rm em}} = \nu_n \frac{1 + z_{\rm em}}{1 + z_{\rm abs}},
\end{equation}
where  $\nu'_n(z_{\rm abs}, z_{\rm em}) < \nu_{n+1}$ (as  higher than $\nu_{n+1}$ frequencies contribute to a cascade from  $\nu_{n+1}$) and  $a = 1/(1+z)$ is the cosmological scale factor. The maximum redshift from which such photons arrive is that of photons emitted just below $\nu_{n+1}$:
 \begin{equation}
    1 + z_{{\rm max},n}(z_{\rm abs}) = (1+z_{\rm abs}) \frac{1-(1+n)^{-2}}{1- n^{-2}},
\end{equation} 
as for the Lyman series $\nu_n \sim 1 - \frac{1}{n^2}$. 

Usually in semi-numerical 21-cm simulations, the photons are assumed to travel in straight lines until they are absorbed, in such a case the distance they reach from their source is:
\begin{equation}
\label{eq:straight_line_distance}
    r(z_{\rm abs}, z_{\rm em}) = \frac{6.0 }{\Omega_m^{1/2} h} \left(\sqrt{a_{\rm abs}} - \sqrt{a_{\rm em}}\right){\rm Gpc}.
\end{equation}
With this assumption, the intensity of \Lya{} photons at a given location $\textbf{x}$ is given by \citep{barkana05}:
\begin{multline}
 \label{eqn:ja_no_r_dist}
    J_{\alpha}(z_{\rm abs}, \textbf{x}) = \sum_{n = 2}^{n_{\rm max}} f_{\mathrm{recycle}}(n) \int_{z_{\rm abs}}^{z_{{\rm max},n}(z_{\rm abs})} \frac{c dz_{\rm em}}{H(z_{\rm em})}  \frac{(1+z_{\rm abs})^2}{4 \pi} \times \\ { \epsilon( \nu'_n(z_{\rm abs}, z_{\rm em}), z_{\rm em} )}_{r(z_{\rm abs}, z_{\rm em}), {\textbf x}}
    \end{multline}
where the photon emissivity, $\epsilon_{r, {\textbf x}}$, is averaged over a spherical shell at a distance $r(z_{\rm abs}, z_{\rm em})$ from ${\textbf x}$ and $r(z_{\rm abs}, z_{\rm em})$ is given by   Eqn. \ref{eq:straight_line_distance}. Until recently, this spherical shell calculation of the radiation fields was the basis of our semi-numerical code \citep[introduced by][]{FialkovLya} which convolves the distribution of sources with a spherical shell window function  in Fourier space. We have revised this prescription taking the effect of multiple scattering of  \Lya{} photons into account when calculating the  \Lya{} intensity field as is described below and in \citep{reis20a}.

\subsection{Multiple scattering of \Lya{} photons}
\label{sec:lya_ms_calculation}
In reality \Lya{}  photons do not travel in straight lines as they  scatter off hydrogen atoms multiple times before being absorbed.   This effect reduces the effective distance at which \Lya{} photons couple the spin temperature to the gas temperature through the WF effect. We have first included this effect in \citet{reis20a} and  summarize the  details of this calculation here for completeness.

Different photons emitted from the same source at the same frequency will scatter a different number of times and at  different scattering angels contributing  to the \Lya{} background at a distribution of effective distances from the source. Distance along a  straight line (Eqn. \ref{eq:straight_line_distance}) is the effective horizon of this  distribution. Out of photons emitted at $z_{\rm em}$ and absorbed at $z_{\rm abs}$, we denote the distance distribution of \Lya{} photons, i. e.  the fraction of photons that travel an effective distance $r$, by $f_{\rm scatter}(z_{\rm abs}, z_{\rm em}, r)$. With this, Eqn. \ref{eqn:ja_no_r_dist} turns to

 \begin{multline} 
 \label{eqn:ja_with_r_dist}
   J_{\alpha}(z_{\rm abs}, \textbf{x}) = \sum_{n = 2}^{n_{\rm max}} f_{\mathrm{recycle}}(n) \int_{z_{\rm abs}}^{z_{{\rm max},n}(z_{\rm abs})} \frac{c dz_{\rm em}}{H(z_{\rm em})}  \frac{(1+z_{\rm abs})^2}{4 \pi} \times \\ \int dr f_{\rm scatter}(z_{\rm abs}, z_{\rm em}, r) \epsilon( \nu'_n(z_{\rm abs}, z_{\rm em}), z_{\rm em})_{r, {\textbf x}},
\end{multline}
where, as before,  $\epsilon_{r, {\textbf x}}$ is averaged over a spherical shell at a distance $r$ from ${\textbf x}$, and, unlike before, we integrate over $r$. 

In the calculation of   $f_{\rm scatter}(z_{\rm abs}, z_{\rm em}, r)$ we employ a  Monte Carlo (MC) code to generate both the distance travelled between two scattering events and the angle of scattering. The code follows the approach taken by \citet{loeb99} with  two modifications: (i) we consider photons absorbed at \Lya{} while \citet{loeb99} considered photons emitted at \Lya{}, (ii) the change of all redshift dependent quantities (such as the Hubble rate) is included exactly (in the high-redshift EdS limit). The  details of the MC procedure are given in Appendix \ref{app:mc}.

\subsection{\Lya{} heating}
\label{sec:lya_heating_intro}

\begin{figure}
    \centering
 \includegraphics[width=0.49\textwidth]{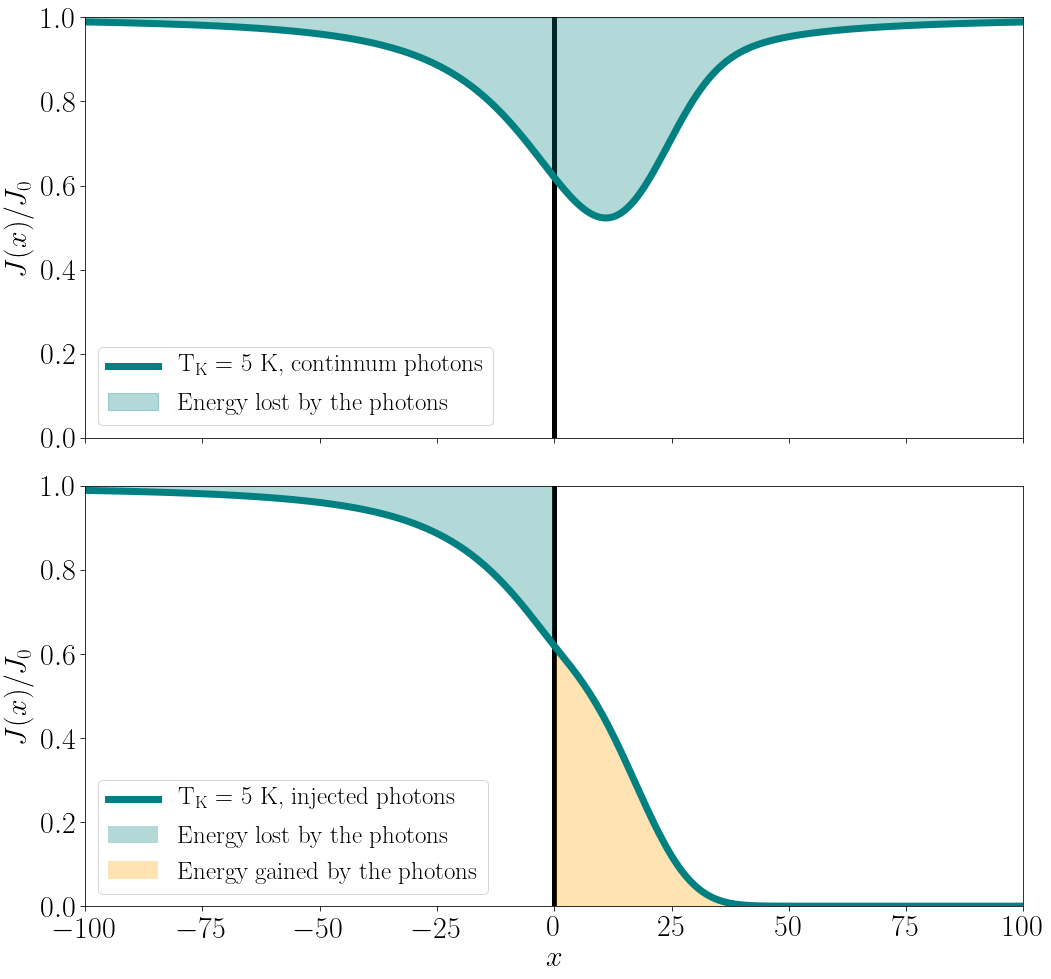}  
    \caption{Spectral shape around the \Lya{} line center for continuum photons (top panel), which are emitted between the \Lya{} and \Lyb{} frequencies and redshifted into the \Lya{} line, and  injected photons (bottom panel), which are produced through atomic cascades by photons emitted between the \Lyb{} and the Lyman limit frequency. The spectral distortion near the line center is due to scattering of the photons with the hydrogen gas. Without scattering, the spectral shape of continuum photons would be $J(x)/J_{0} = 1$, and of injected photons  $J(x)/J_{0} = 1 - \Theta(x)$. The difference between the original and distorted shapes measures the energy transfer between the photons and the  gas.}
    \label{fig:Jx}
\end{figure}

The same stellar photons that trigger \Lya{} coupling also contribute to  heating of the IGM through  recoils of the gas atoms as a result of scattering.
Heating rate due to \Lya{} photons depends on the balance between  injected and continuum photons. This is because,  as first pointed out by \citet{chen04}, injected photons cool the gas, while continuum photons heat the gas (we discuss this point in detail below, Section \ref{sec:heating_rate_calc} and Fig. \ref{fig:Jx}). Historically, the impact of \Lya{} photons on the thermal state of the IGM  was thought to be negligible  owing to the assumption that for a standard intrinsic source  spectrum the number of photons emitted between \Lya{} and \Lyb{} frequencies (which redshift into the \Lya{} line, and thus constitute the continuum radiation) is very close to the number of photons emitted between the \Lyb{} and Lyman limit frequencies (which are the source of the injected photons). With this assumption   heating is nearly fully compensated by cooling  \citep{chen04}. However, as noted in later works \citep{furlanetto06b, hirata06, chuzhoy07}, the numbers of injected and continuum photons are not similar, with substantially more continuum photons expected. This is due to the fact that most photons emitted between the \Lyb{} and Lyman limit frequencies will cascade into the 2s level and produce two photons with frequencies smaller than \Lya{}. These photons, therefore, are removed from the pool and do not contribute to cooling. Only a fraction of these photons cascades into the 2p level and, subsequently, produces injected \Lya{} photons. Given the expected ratio between  continuum and injected photons, \Lya{}  can heat the gas to a temperature of $\sim 100$ K  \citep{chuzhoy06}. At this temperature \Lya{} effect is self-regulated and cooling  annuls heating. That is, \Lya{} photons cannot heat the gas to higher temperatures, and, if the gas is heated  by some other  mechanism (e.g. by X-rays), \Lya{} photons will act to cool the gas. 

For completeness, we note that \citet{meiksin20} have recently considered a significantly different setup in which \Lya{} photons can cool the gas producing a deep 21-cm absorption feature as found by EDGES \citep{bowman18}.  While all of the previously mentioned  works (including this paper) consider  \Lya{} photons resulting from stellar continuum,  \citet{meiksin20} calculate the effect of \Lya{} photons originating directly from line emission (for example from recombination in HII regions) assuming extreme  \Lya{} line emission strength and profile. This  contribution is almost always neglected in 21-cm studies, and its importance is unclear, not just due to the sensitivity to complex local effects as studied by \citet{meiksin20}, but also because of a doubtful assumption in their calculation. They did not account for the \Lya{} scattering that we include here, because they assumed that the scale of this scattering is large compared to the typical spacing between \Lya{}-emitting galactic halos, so that the sources can be treated as spatially uniform. As we find here and in our other recent publications, at cosmic dawn the \Lya{} field is in fact highly clustered around the (rare) sources and is highly non-uniform. Also, within each \Lya{} bubble surrounding a source, the majority of \Lya{} scatterings occur close to the center, on a significantly smaller spatial scale than the scale at which the \Lya{} photons escape (which is the smoothing scale that they assumed).

\subsubsection{Calculation of the \Lya{} heating rate}
\label{sec:heating_rate_calc}

In the calculation of the \Lya{} heating rate we closely follow \citet{chuzhoy07} except for relaxing an approximation (see below). 
Spectral shape in the vicinity of the \Lya{} line   is given by \citep{grachev89, chuzhoy06, chuzhoy07}
\begin{equation}
    J(x) = 2 \pi J_{0} \gamma a^{-1} \int_{-\infty}^{x} e^{\frac{-2 \pi \gamma (x^3 - z^3)}{3a} - 2\eta (x-z)} z^2 dz
    \label{eq:Jxc}
\end{equation}
for   continuum photons and for the red wing (x < 0) of  injected  photons;
while the  blue wing (x > 0) of the profile for the injected photons can be written as
\begin{equation}
    J(x) = J(0) e^{\frac{-2 \pi \gamma x^3}{3a} - 2\eta x},
    \label{eq:Jxi}
\end{equation}
where  $a = A_{21}/4 \pi \Delta \nu_{\rm D}$ and $A_{21}$  is the spontaneous decay rate of the \Lya{} transition, $\Delta \nu_{\rm D}$ is the Doppler width of the line; $\gamma = 1/\left[\left(1+ 0.4/T_{\rm S}\right) \tau_{\rm GP}\right]$ where $\tau_{\rm GP}$ is the Gunn Peterson optical depth; $\eta = h \nu_{\alpha}/\sqrt{2 k_{\rm B} T_{\rm K}m_{\rm H} c^2} \left(1+ 0.4/T_{\rm S}\right)\left(1+ 0.4/T_{\rm K}\right)^{-1}$ where $h$ is the Planck constant and $\nu_{\alpha}$ is the \Lya{} frequency; $J_{0}$ is the radiation intensity far away from the line center, unaffected by scattering, and $J(0)$ is the intensity at $x = 0$. The parameter $x$ is defined as 
\begin{equation}
    x = \left(\frac{\nu}{\nu_{\alpha}} -1\right) / \sqrt{\frac{2 k_{\rm B} T_{\rm K}}{m_{\rm H} c^2}} = \left(\frac{\nu}{\nu_{\alpha}} -1\right)  \frac{\nu_{\alpha}}{\Delta \nu_{\rm D}} = \frac{\nu - \nu_{\alpha}}{\Delta \nu_{\rm D}},
\end{equation}
where  $m_{\rm H}$ is the hydrogen mass.

 \citet{furlanetto06b} compared the approximated spectral shape given by  Eqs. \ref{eq:Jxc} and \ref{eq:Jxi} to more detailed numerical calculations, and found that the approximation is excellent for $T_{\rm K} < 1000$ K. This covers the entire region of interest, since, as we mentioned above, \Lya{} photons can only heat the gas up to $T_{\rm K} \sim 100$ K.

Examples of the  spectral shape around the \Lya{} line center for continuum and  injected photons are shown in Fig. \ref{fig:Jx} for the specific case of $z = 15$, $T_{\rm K} = 5$ K and  $T_{\rm S} = T_{\rm K}$.  For continuum photons (top panel of Fig. \ref{fig:Jx}) the absorption feature near the line center represents energy lost by the photons and transferred to the gas. Thus, the gas atoms gain energy and their kinetic temperature rises. For injected photons, the intrinsic spectral shape can be written as  $J(x)/J_{0} = 1 - \Theta(x)$ where $\Theta(x)$ is the  Heaviside step function, i.e. without scattering $J(x) = 0$ on the blue side ($x>0$) and $J(x)/J_{0}=1$ on the red side ($x<0$) of the  \Lya{}  line center. With scattering included, the intensity seen on the red side  goes into heating of the IGM, while the feature on the blue side  represents energy gained by the photons and lost by the gas. Thus, in total, injected photons can cool the hydrogen gas.

Energy gained by the gas per  \Lya{}  photon can be calculated as   
\begin{equation}
    \Delta E = h \Delta \nu_{\rm D}  \int  dx \left(1 - \frac{J(x)}{J_{0}} \right), 
\end{equation}
for continuum photons, and 
\begin{equation}
    \Delta E = h \Delta \nu_{\rm D}  \int  dx \left(1 - \Theta(x) - \frac{J(x)}{J_{0}} \right)
\end{equation}
for injected photons \citep{chen04}. The energy gain depends on the kinetic temperature of the gas, the spin temperature and the redshift. We perform the full calculation of the energy gain  on a grid of these properties and use these results in the calculation of the thermal history. We verify that our results agree well with the approximations from \citet{chuzhoy07} and \citet{furlanetto06b} when the kinetic temperature is not small and these approximations are valid. 
Finally, the  \Lya{}  heating rate  per hydrogen atom (in erg s$^{-1}$) is given by
\begin{equation}
\label{eqn:lya_heating_rate}
 \epsilon_{\rm Ly_{\alpha}} = \dot{N}_{\alpha}^{\,\rm continuum} \Delta E^{\,\rm continuum}  + \dot{N}_{\alpha}^{\,\rm injected} \Delta E^{\,\rm injected},
\end{equation}
where $\dot{N}_{\alpha}$ is the number of photons reaching the line center per hydrogen atom per unit time 
\begin{equation}
    \dot{N}_{\alpha}^{\,\rm continuum, \; injected}  = \frac{4 \pi \nu_{\alpha} dz/dt}{n_{\rm H} (1+z) c  } J_{\alpha}^{\,\rm continuum, \; injected}, 
\end{equation}
where $J_{\alpha}^{\,\rm continuum, \; injected}$ (in units of cm$^{-2}$ sr$^{-1}$ s$^{-1}$ Hz$^{-1}$) is calculated in our simulation as described in \ref{sec:lya_ms_calculation} (we separate the contributions of injected and continuum photons).

\section{Results}
\label{sec:results}

\subsection{Effect of multiple scattering of \Lya{} photons on the 21-cm signal}
\label{sec:lya}

\begin{figure}
    \centering
 \includegraphics[width=0.99\columnwidth]{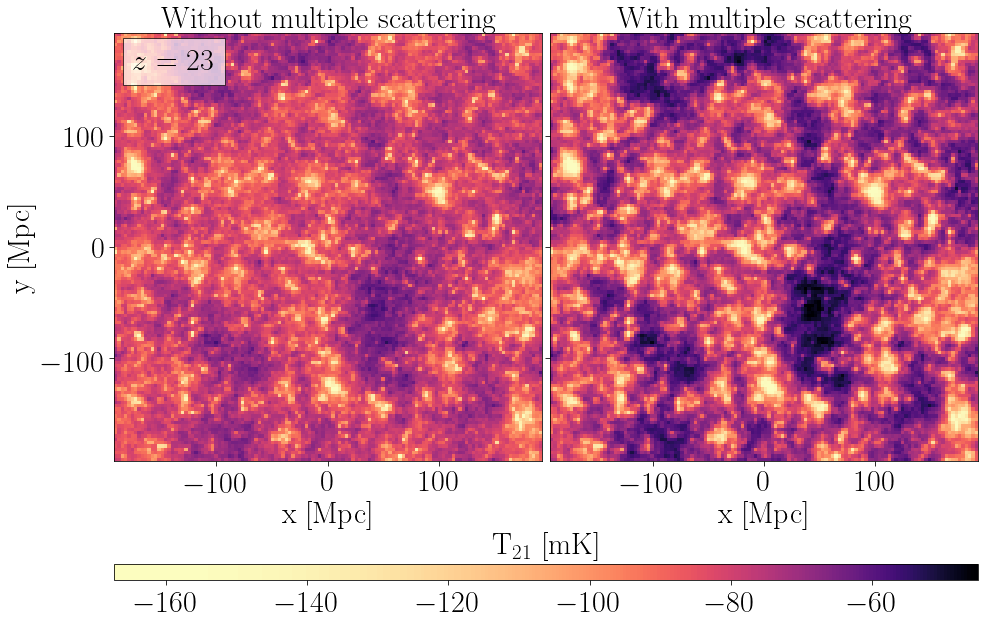} 
 \includegraphics[width=0.99\columnwidth]{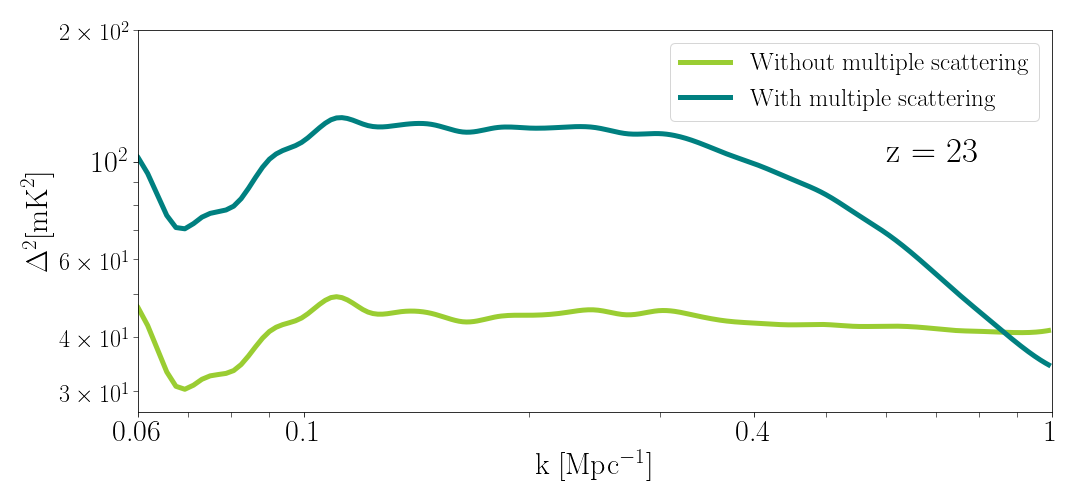}

    \caption{ The effect of multiple scattering on 21-cm fluctuations. {\bf Top:  }  A slice of the 21-cm brightness temperature from  the era of \Lya{} coupling (shown at $z=23$), with ({\bf right}) and without ({\bf left}) the effect of multiple scattering of \Lya{} photons on the WF coupling. This example shows a simulation  with $V_{c} = 16.5$ km s$^{-1}$, $f_{\ast} = 0.05$,  $f_{X} = 1$ and a hard X-ray SED of XRBs, with \Lya{} and CMB heating included.  {\bf Bottom:  } Corresponding power spectra at $z=23$ for the full case with multiple scattering (dark green) and for the case without multiple scattering (light green). 
     } 
    \label{fig:t21_map}
\end{figure}

Through its impact on the coupling coefficient $x_\alpha$, multiple scattering leads to redistribution of power in the 21-cm fluctuations.
In this section we explore this effect comparing simulation results  with and without multiple scattering. The only difference between the two types of simulations is in how  the \Lya{} background is calculated. In the former case the new window functions (as described in \cref{sec:lya_ms_calculation}) are utilized, while in the latter we assume that photons propagate along straight lines and use spherical shells window functions. Example images of the 21-cm brightness temperature from cosmic dawn,  with and without the effect of multiple scattering,  are shown in the top panels of Fig. \ref{fig:t21_map}. The simulations assume  $V_{c} = 16.5$ km s$^{-1}$,  $f_{\ast} = 0.05$,   $f_{X} = 1$ and hard X-ray SED of XRBs.  \Lya{} and CMB heating are included in the simulations discussed in this Section.

 With this effect included, the strength of the coupling is suppressed in regions far away from the radiation sources and enhanced in regions closer to the  sources leading to weaker 21-cm absorption in distant areas and stronger signal (in absorption) closer to the sources. This behavior is a natural outcome of multiple scattering, with which \Lya{} photons travel shorter effective distances from the sources before being absorbed, and leads to an enhanced large-scale 21-cm power spectrum  (as shown in the bottom panel of Fig. \ref{fig:t21_map}).  In addition, multiple scattering washes out fluctuations at distances shorter than the mean free path, thus suppressing 21-cm power spectrum on small scales.  In terms of timing,  in regions around the radiation sources there will be more \Lya{} photons, and the \Lya{} coupling transition will occur sooner compared to more isolated areas. These effects are clearly seen on the images (Fig. \ref{fig:t21_map}) where the purple/black regions of  weak 21-cm signal  are the ones far away from the radiation sources, and there the signal gets even weaker  due to multiple scattering.

\begin{figure}
    \centering

 \includegraphics[width=0.99\columnwidth]{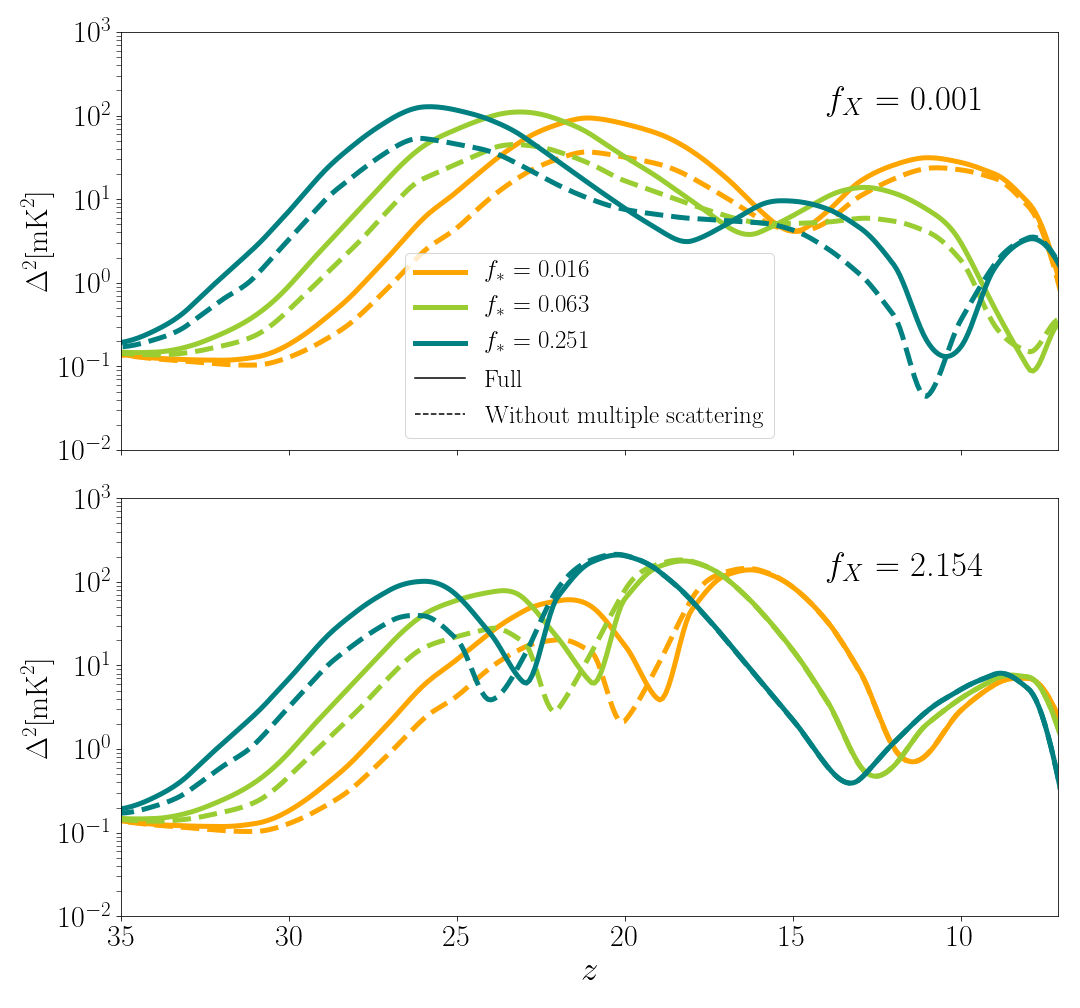} 
    \caption{The 21-cm power spectra from simulations with and without the effect of multiple scattering. We assume $V_{c} = 16.5$ km $^{-1}$ and  a soft  X-ray power law SED with minimum energy $E_{\rm min} = 0.1$ keV and $\alpha$ = 1.5. The results are shown for two different values of $f_X$, 0.001 ({\bf top}) and 2.154 ({\bf bottom}), and different values of the star formation efficiency $f_{\ast}$ (as indicated in the legend).  For each value of $f_{X}$ and $f_{\ast}$ the solid/dashed line shows the power spectrum with/without the effect of multiple scattering.  The power spectra are shown as a function of redshift, for $k = 0.1$ Mpc$^{-1}$. An increase in the strength of the power spectra during the era of \Lya{} coupling is seen in all cases. For cases with weak X-rays, in which \Lya{} heating is the dominant heating mechanism, an increase is seen also during the heating era, which occurs rather late in these models (these results are discussed in detail in Section \ref{sec:lya_cmb_heating_res}). }
    \label{fig:power_same_vc_diff_fs}
\end{figure}

Out of the  effects listed above, perhaps the most interesting one   is  the enhancement  of the 21-cm fluctuations on large   observable scales. To further explore this effect, we show (in Fig. \ref{fig:power_same_vc_diff_fs})  the redshift evolution of the 21-cm power spectrum  at $k =$  0.1 Mpc$^{-1}$ with and without multiple scattering for several different astrophysical scenarios. At these scale we see a significant amplification of the power spectrum during the \Lya{} coupling era for all demonstrated astrophysical scenarios. This suggests that the inhomogeneous coupling field is the dominant source of fluctuations at this value of $k$ and corresponding high redshifts. At lower redshifts, where \Lya{} coupling saturates and the 21-cm fluctuations are dominated by heating processes, the effect of \Lya{} multiple scattering becomes irrelevant in most cases. However, in scenarios with weak X-ray heating the effect of multiple scattering can be noticeable even at low redshifts. In such models \Lya{} heating is a dominant heating mechanism (see further discussion  in Section \ref{sec:lya_cmb_heating_res}) and as such, the spatial distribution of \Lya{} photons remains important during the heating era. Even though \Lya{} heating requires strong \Lya{} background and, thus, is very smooth (see Section \ref{sec:lya_cmb_heating_res}), we do find small intrinsic non-homogeneity, as discussed in Appendix  \ref{app:PS}.

\begin{figure}
    \centering
 \includegraphics[width=0.99\columnwidth]{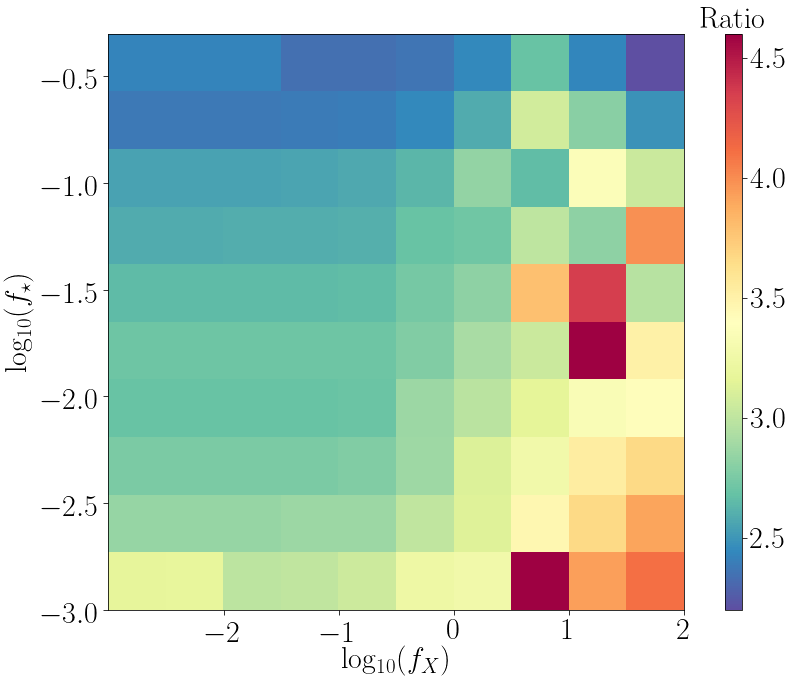}  
    \caption{Impact of multiple scattering on the 21-cm power spectrum (colorbar on the right, calculated according to Eqn. \ref{eqn:normalized_diff})  calculated for different values of $f_X$ (x-axis) and $f_*$ (y-axis). The effect is  calculated at   $k$ = 0.1 Mpc$^{-1}$. We use $V_c$ = 16.5 km s$^{-1}$ and  a soft  X-ray power law SED with minimum energy $E_{\rm min} = 0.1$ keV and $\alpha$ = 1.5 (same as Fig. \ref{fig:power_same_vc_diff_fs}).}
    \label{fig:ratio}
\end{figure}

At high redshifts at which multiple scattering plays an important role the main astrophysical process  that affects the 21-cm signal is that of star formation, which in our simulation is parametersized by $f_*$ and $V_c$. In addition,  Fig. \ref{fig:power_same_vc_diff_fs} indicates  that   multiple scattering can be important during the heating era in cases with low $f_X$ and/or hard SED, while it plays a subdominant role during reionization. Therefore, in order to assess the importance of multiple scattering over the entire viable astrophysical space, we perform a parameter study varying $f_*$ and $f_X$   while keeping the rest of the model parameters fixed. We quantify the strength of the effect of multiple scattering by calculating the maximum ratio between the peak value of the 21-cm power spectrum in mK$^2$ units, $\Delta^2$, in models with the effect included  and the corresponding case without it,
\begin{equation}
\label{eqn:normalized_diff}
    {\rm Maximum\; ratio} = {\rm Max}\left(\frac{\Delta^2_{\rm With \; MS} }{\Delta^2_{\rm Without \; MS}}\right),
\end{equation}
where the maximum is taken over the wide range of redshifts from $z = 35$ to $z = 6$. We calculate the ratio at  $k = 0.1$ Mpc$^{-1}$  and show the results in Fig. \ref{fig:ratio}. We find  a factor of 2-5 enhancement, with the largest effect being in the  case of late and inefficient star formation but high X-ray production efficiency. We  also verify that the effect is robust with respect to $V_c$. For example, for the fixed value of $f_*$ = 0.03 and $f_X$ = 1 the maximum ratio change from  2.7 to 2.1 when varying $V_c$ from 16.5 km s$^{-1}$ to 76.6 km s$^{-1}$ .

Finally, we note that multiple scattering slightly reduces the global 21-cm signal.   While the total number of  \Lya{} photons is conserved, the efficiency of the \Lya{} coupling is changed due to scattering. This is because \Lya{} photons are more concentrated in  regions around the sources, and more of them can be effectively "wasted" (in terms of their contribution to the WF coupling) in areas where the coupling is already saturated. On the other hand, in more isolated regions the number of  \Lya{} photons is lower and, thus, the WF coupling is less efficient, which  leads to a weaker 21-cm signature. For the cases shown in Fig. \ref{fig:power_same_vc_diff_fs} the maximum difference in the global signal ranges from $\Delta T_{21} = 2.5 $ mK ($\sim 1.5 \%$ of the maximum depth of the signal) for $f_{\ast} = 0.5$ to $\Delta T_{21} = 0.5$ mK ($\sim 0.5 \%$ of  maximum depth) for $f_{\ast} = 0.001$.

\subsection{The effect of \Lya{} and CMB heating}
\label{sec:lya_cmb_heating_res}

\begin{figure*}
    \centering
  \includegraphics[width=0.99\textwidth]{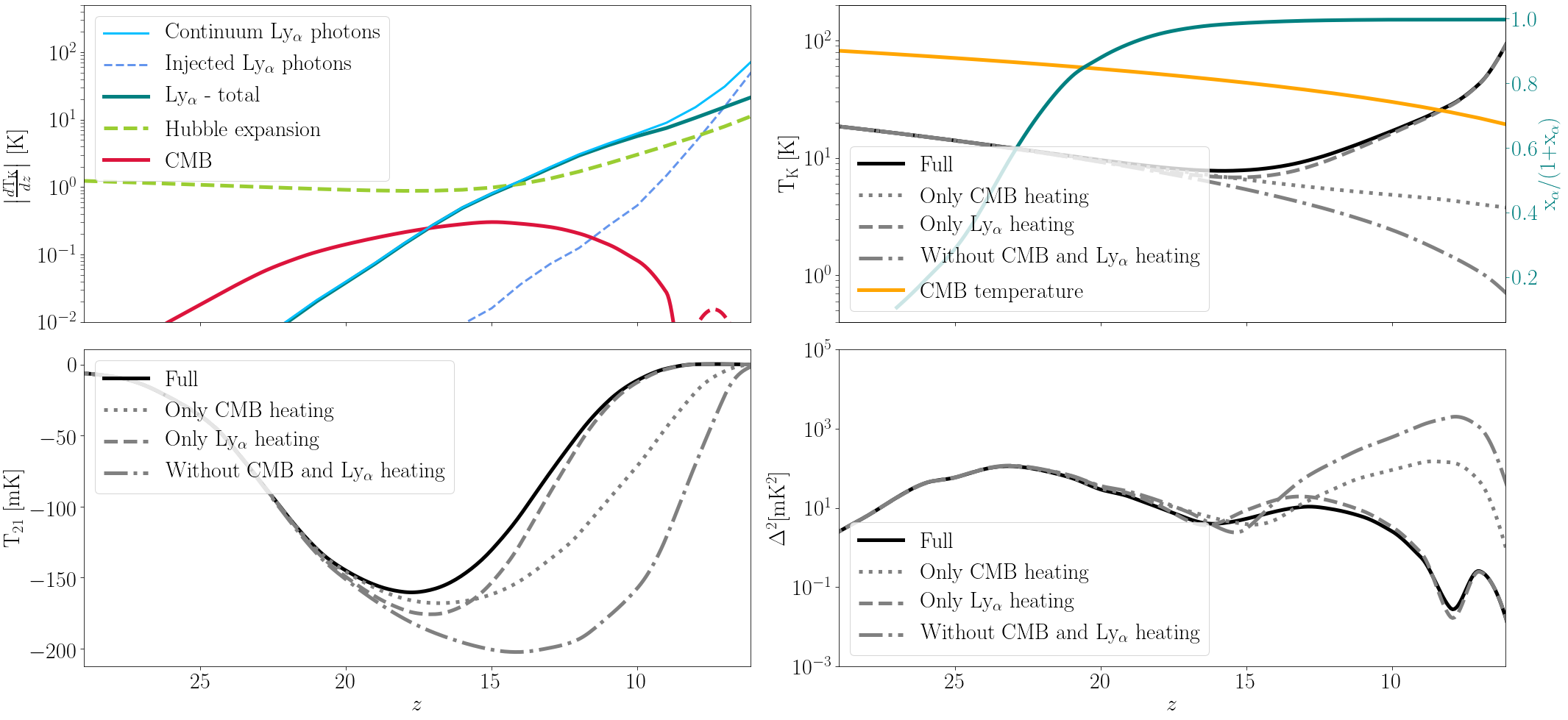}  \\ [6pt]
    \caption{
Thermal history (averaged over the simulation box) and the resulting 21-cm signal for an  example astrophysical scenario where heating is dominated by \Lya{} photons. We assume $f_X = $ 0 (no X-rays), $f_*$ = 0.05, $V_c$ = 16.5 km s$^{-1}$.   {\bf Top left:} Heating/cooling rates of the neutral IGM  (see the legend). Solid lines show mechanisms that heat the IGM (negative $d{\rm T}_{\rm K}/dz$) and dashed lines show mechanisms that cool the IGM (positive $d{\rm T}_{\rm K}/dz$).  In the interest of space, we do not show the contributions of Compton heating and structure formation which are included in our simulation. {\bf Top right:} The gas temperature with and without \Lya{} and CMB heating (see the legend). We also show the evolution of the coupling coefficient (solid blue) to illustrate that \Lya{} heating becomes important only after the coupling nearly saturates. {\bf Bottom left:} The global 21-cm signal with and without \Lya{} and CMB heating (see the legend).  {\bf Bottom right:} The 21-cm power spectrum at $k = 0.1$ Mpc$^{-1}$ with and without \Lya{} and CMB heating (see the legend).  } 

    \label{fig:heating_rates}
\end{figure*}

We now turn to consider the effects of  \Lya{}  and CMB heating on the 21-cm signal for models with no X-ray heating (Fig.  \ref{fig:heating_rates}).  
While the effect of \Lya{}  heating on the global 21-cm signal was estimated for a few selected astrophysical models  \citep[e.g.][]{ciardi07, ciardi10, mittal20}, implications for  the 21-cm fluctuations are explored in this paper for the first time. In addition, with the onset of star formation and buildup of  \Lya{}  background, both  \Lya{}  and CMB heating channels are enabled and should be taken into account simultaneously. Effects of both processes on the global 21-cm signal (but not on the fluctuations) were considered by \citep{venumadhav18}; however, in their calculation of  \Lya{}  heating the authors followed \citet{chen04} thus underestimating this effect.

Evolution of  \Lya{} and CMB heating and cooling rates for a model with $f_* = 0.05$, $V_c = 16.5$ km s$^{-1}$ and $f_X=0$ is shown in the top left panel of Fig. \ref{fig:heating_rates}, along with the cooling rate due to the expansion of the Universe  for comparison\footnote{Note that in the interest of space we are not showing components such as Compton heating and shock heating due to structure formation which are included in our simulation.}. Fig. \ref{fig:heating_rates} also shows the redshift evolution of gas kinetic temperature as well as the global 21-cm signal and the power spectrum. Additional examples with $f_* = 0.01$,  $V_c = 16.5$ km s$^{-1}$ and $f_* = 0.2$, $V_c = 50$ km s$^{-1}$ can be found in Appendix \ref{app:heating_rates}. 

From the figure we see that, despite being sub-dominant over a wide redshift range, the CMB heating rate does have an effect on the overall thermal history and the resulting 21-cm  signal. Even though it is generally weak (and is always weaker than cooling by Hubble expansion), CMB heating is  stronger than \Lya{} heating at redshifts higher than $z\sim 17$ (for the model shown in the figure) and contributes a few degrees to the gas temperature (top right panel of Fig.  \ref{fig:heating_rates}),  which in turn reduces  the depth of the global 21-cm signal by a few tens of mK and suppresses the power spectrum (bottom panels of Fig.  \ref{fig:heating_rates}).
 
The effect of \Lya{} heating starts to dominate at $z\sim 17$, first over the CMB heating and then over the Hubble expansion term. The leading effect is that of continuum photons which heat the gas, with the contribution of injected photons  becoming important only at low redshifts following the increase  of the kinetic gas temperature. In this specific example, with the  \Lya{}  and CMB heating included, gas temperature  reaches  the  value of $ 101.2$ K  by the end of the simulation ($z=6$, the lowest redshift shown); while when neglecting the two contributions the temperature at $z = 6$ is only $0.6$ K, and when only  CMB heating is taken into account it is $3.7$ K.
Because in the full case (with both \Lya{}  and CMB heating included) the temperature is always lower than the threshold value at which heating by continuum \Lya{} photons is canceled by cooling due to the injected photons,  \Lya{} photons keep heating up the neutral gas even at the end of reionization.

Additional intuition regarding the redshift range at which \Lya{} heating is significant can be obtained by considering the coupling coefficient (shown with the blue line in the top right panel of Fig. \ref{fig:heating_rates}). Comparing the time dependence of the coupling coefficient to that of the gas temperature  in the model with \Lya{} heating, we notice that \Lya{} heating does not have an impact on thermal history for $x_\alpha \ll 1$, i.e. when the  \Lya{} background is weak. Efficient  \Lya{} heating requires a strong intensity of the \Lya{} field and, thus, is expected to affect the low-redshift 21-cm signal while having a minor effect at high redshifts. 

\begin{figure}
    \centering
 \includegraphics[width=0.99\columnwidth]{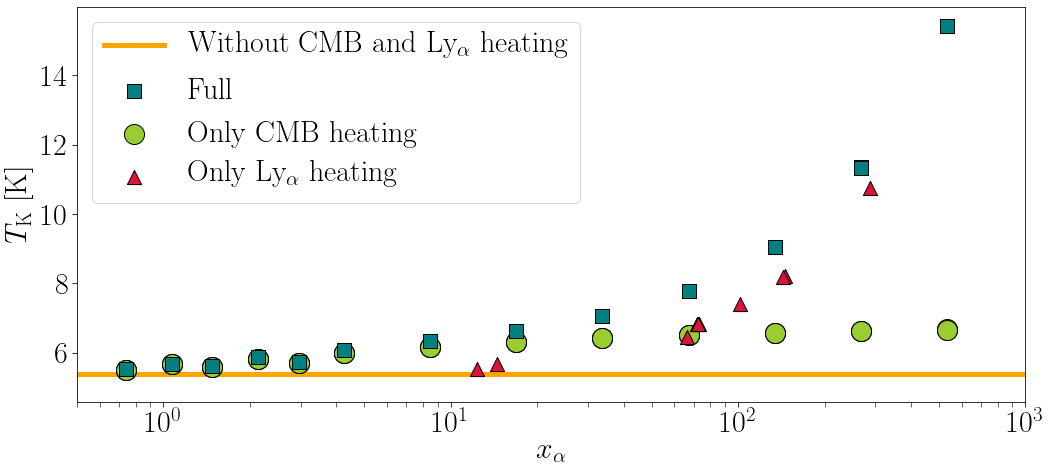}  
    \caption{The kinetic gas temperature vs.\ the \Lya{} coupling coefficient at $z = 15$. Each point represents a simulation run with different astrophysical parameters. Simulation runs with both \Lya{} and CMB heating are shown with dark green squares and compared to simulations with only CMB heating (light green circles) or only \Lya{} heating (red triangles). The horizontal line marks the reference case of an adiabatically cooled gas with no \Lya{} heating and no CMB heating.}  
    \label{fig:tk_vs_xa}
\end{figure}

That significant \Lya{} heating occurs following a significant delay after the onset of WF coupling was first pointed out  by \citet{furlanetto06b}, where an approximation for the fractional temperature change per Hubble time from \Lya{} heating was derived for small values of $x_{\alpha}$. Their result,
\begin{equation}
    \frac{2}{3}
    \frac{        \epsilon_{\rm Ly_{\alpha}}   }
    {H(z) n_{\rm HI} k_{\rm B} T_{\rm K} } \sim
    x_{\alpha} \frac{0.8}{  T_{\rm K}^{4/3}   } \frac{10}{1+z},   
\end{equation}
valid for the early stages of \Lya{} heating, suggests  that \Lya{} heating  is negligible when $x_{\alpha} \ll 1$,  and the 21-cm spin temperature deviates from the CMB temperature only when  $x_{\alpha}\gtrsim 1$. Our results agree with this conclusion, as can be seen from Fig. \ref{fig:tk_vs_xa} which shows the kinetic temperature versus the \Lya{} coupling coefficient at a fixed redshift of $z = 15$ for models with \Lya{} plus CMB heating, CMB heating only and a reference case without the two terms. The results are shown for a large selection of  astrophysical scenarios with $f_*$ varied between 0.001 and 0.5, $V_c$ between 16.5 km s$^{-1}$ and 76.5 km s$^{-1}$, and $f_X=0$ for all the cases.  
The two sets with heating start to diverge from one another at $x_{\alpha} \sim 10$, showing the  the effect of \Lya{} heating is negligible at lower $x_{\alpha}$, but wins over the CMB heating at higher values of the coupling coefficient.  An important implication is that the commonly used limit of fully coupled and adiabatically cooled gas, even though an insigtful approximation,  is, actually, not a valid physical case. This is because  strong \Lya{} coupling will inevitably be accompanied by substantial   \Lya{} and CMB heating which significantly change the thermal state of the gas compared to the adiabatic cooling limit. With the \Lya{} and CMB heating taken into account, gas temperature in the absence of X-ray heating is model-dependent and varies as a function of $f_*$ and $V_c$. We leave detailed characterization of this limit for future work.

Let us now examine the resulting 21-cm signals in greater details (bottom  panels of Fig. \ref{fig:heating_rates}).   As is evident from the figure, both \Lya{} and CMB heating  play important roles in shaping the  absorption trough of the global signal. Because at high redshifts  CMB heating rate is higher than that of \Lya{}, it has a stronger effect on the depth of the absorption feature, while \Lya{} heating plays a more important role in shaping the low-redshift part of the global signal. We find that in the full case the maximum depth of the absorption feature is only $-160.1$ mK, i.e. the absorption feature is $\sim 20\% $ shallower than when the extra heating terms are not included ($-202.4$ mK). For comparison,  when only CMB heating is considered, the maximum depth is $-167.9$ mK and when only \Lya{} heating is considered the maximum depth is $-175.8$ mK. In total, with the extra heating terms included, the absorption feature  is shallower, narrower and  shifted to lower redshifts. In addition, a weak emission feature might be present  if the gas is heated above the temperature of the CMB.

Another important implication of the extra heating terms is the suppression of  the 21-cm power spectrum  at low redshifts where their contribution is significant  (at $z \lesssim 15$ in the specific example shown in the  bottom right panel of Fig. \ref{fig:heating_rates}). Because \Lya{} heating by individual sources is generally weak,  this component becomes relevant only at low redshift after a significant population of sources builds up to produce strong enough \Lya{} background (as we discussed above, $x_\alpha\gg1$ is required). Therefore, heating by  \Lya{} photons is relatively smooth (see a detailed discussion on the intrinsic non-homogeneity of   \Lya{} contribution to heating in Appendix \ref{app:PS}) and its effect  on the 21-cm power spectrum is mostly a direct manifestation of the suppression of the total signal intensity: since the total magnitude of the absorption 21-cm signal is reduced due to the additional heating by  \Lya{} and CMB, so is the strength of the power spectrum. 

The change in the power spectrum is most strongly manifested around the redshift of heating transition, $z_h$, defined as the moment when the temperature of the gas is equal to the radiation temperature. At $z_h$ the resulting global 21-cm signal vanishes leading to  vanishing power in fluctuations. With the extra heating terms included, this milestone is  shifted to earlier cosmic times (higher redshifts) and the power spectrum at $z_h$ is stronger than expected. Specifically for the scenarios with no X-ray heating that we consider here, the 21-cm signal is strong in absorption during the epoch of reionization and is manifested as a  prominent peak in the power spectrum when shown at a fixed comoving scale and as a function of redshift. Inclusion of either one of the new heating terms (\Lya{} and/or CMB) decreases the contrast between the temperature of the gas and that of the CMB,  shifting the point of heating transition to higher redshifts, which  leads to elimination of the low-redshift power peak. 

In total, we find that suppression of the power spectrum  during reionzation,  shallower  and narrower global absorption trough and emergence of a (weak) emission feature  are  generic implications of \Lya{} and CMB heating and have important consequences for the 21-cm signals in the observable redshift range, as further discussed in \cref{sec:ranges}.

\subsection{\Lya{} and CMB heating versus X-ray heating}

As we mentioned above, \Lya{} photons alone can heat the gas up to a temperature of $\sim 100$ K.  While efficient X-ray heating can bring atoms to much higher kinetic temperatures, the actual X-ray production capability of high redshift galaxies is not well constrained and could be very low. As we have just seen, in such cases   \Lya{} and CMB heating are expected to have a strong effect on the predicted 21-cm signal. In this section we explore the interplay between  X-ray heating and the joined contribution of   \Lya{} plus CMB heating, in order to find values of astrophysical parameters for which \Lya{} and CMB terms should not be ignored.

\begin{figure}
    \centering
 \includegraphics[width=0.99\columnwidth]{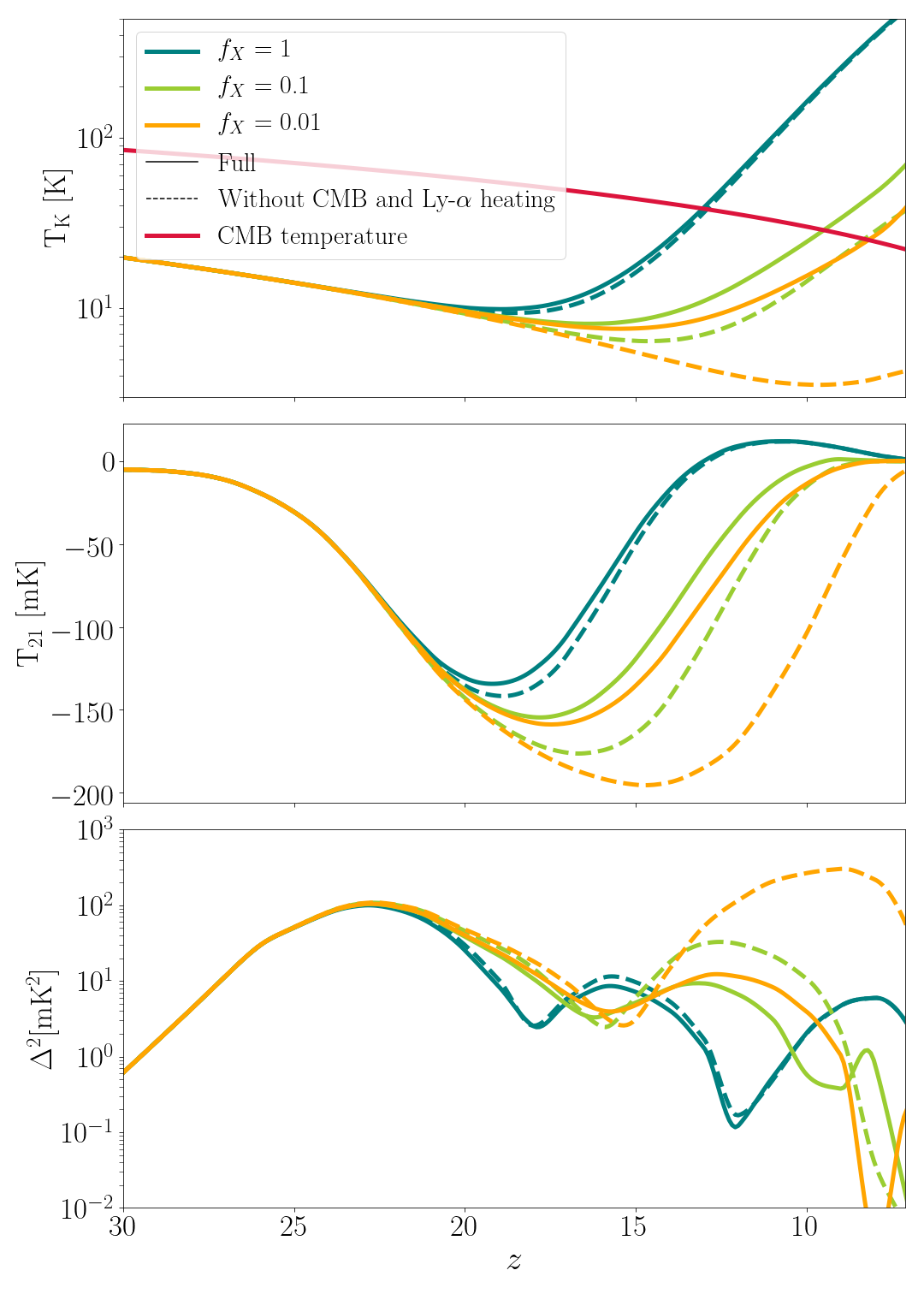}  
    \caption{The effect of \Lya{} and CMB heating in models with various levels of X-ray heating. {\bf Top panel:} The effect of \Lya{} and CMB heating on the kinetic temperature of the gas for various values of the X-ray production efficiency (solid lines). A comparison model, where \Lya{} and CMB heating are both not present, is also shown (dashed).  {\bf Middle panel:}  The global 21-cm signal for the same astrophysical cases. {\bf Bottom panel:} The  power spectrum at $k$ = 0.1 Mpc$^{-1}$. In this figure we use the same set of astrophysical parameters as in Fig. \ref{fig:heating_rates} (except for varying $f_X$ as is indicated in the legend).} 
    \label{fig:heating_global_ps}
\end{figure}

We start by considering the effect of \Lya{} plus CMB heating on the gas kinetic temperature, the 21-cm global signal and power spectrum for  three different values of  the X-ray production efficiency $f_X = 0.01$, 0.1, 1 and  hard SED of XRBs (Fig. \ref{fig:heating_global_ps}). We use the same base model as in Fig. \ref{fig:heating_rates} with $f_* = 0.05$ and  $V_c = 16.5$ km s$^{-1}$. We find that for low (but non-negligible) values of $f_X$ (namely, 0.01 and 0.1),  \Lya{} plus CMB heating is the dominant heating mechanism, and, consequently,  the effect of the extra heating terms on the 21-cm signal is very significant. When including \Lya{} and CMB heating, similarly to what we saw in the case with $f_X=0$, with weak X-ray heating  the absorption trough of  the global signal becomes narrower, shallower, and with a minimum at higher redshift, while the low-redshift power spectrum is strongly suppressed. A minor effect  is seen even in the case of $f_X =1$, i.e. the most commonly used  X-ray heating efficiency.

Like in the case with no X-ray heating, in the presence of  X-ray sources the change in the power spectrum is most strongly manifested around the heating transition. For example, in the case with $f_X=0.01$ and without  \Lya{}  and CMB heating terms, even though it is warmer than in the adiabatic cooling limit, the gas is still colder than the CMB at all the simulated redshifts. However,  with the extra heating terms included  the point of heating transition is shifted to higher redshifts ($z\sim 7$), and the EoR power spectrum is strongly suppressed. More generally, in models with weak  X-ray heating which usually leads to cold  reionization (i.e. the IGM is colder than the CMB during the EoR), extra heating acts to suppresses 21-cm fluctuations by reducing the contrast between the background temperature and that of the gas;  while in hot EoR scenarios (where  the IGM is heated above the CMB level early on), extra heating will have a negligible effect as CMB heating is small compared to X-rays while \Lya{}  heating vanishes owing to the cancellation of the continuum and injected contributions. 

\begin{figure}
    \centering
 \includegraphics[width=0.99\columnwidth]{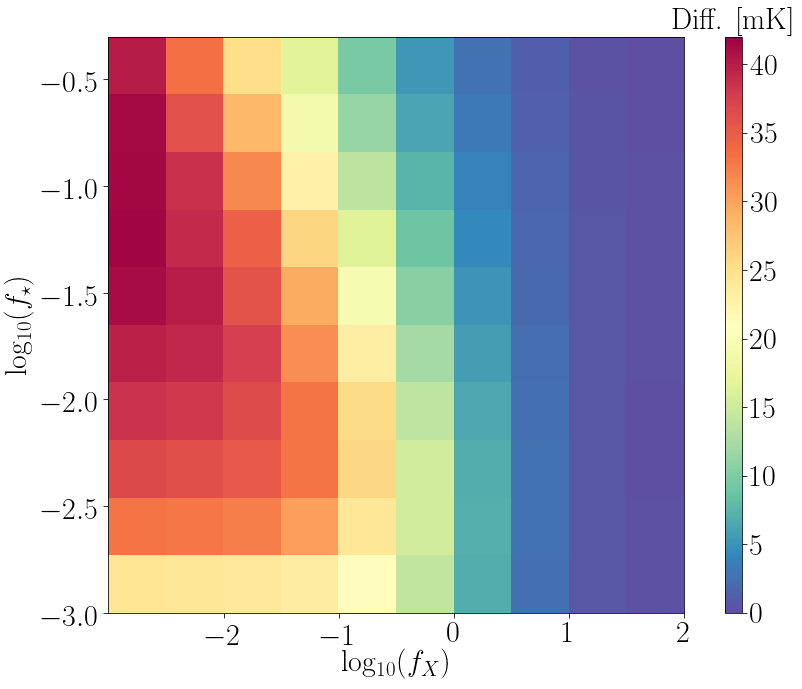}  
    \caption{Difference (in mK units) in the intensity of the global absorption feature between simulations with \Lya{} and CMB heating and reference simulations  without these two effects. The difference is shown as a function of $f_X$ and $f_*$ for a fixed value of  $V_c  = 16.5$ km s$^{-1}$ and assuming the predicted high-$z$ SED of XRBs. }
  
    \label{fig:heating_effect_size}
\end{figure}

Finally, we explore a grid of models in order to establish at which values of astrophysical parameters  \Lya{}  and CMB heating  are important relatively to X-ray heating.   Since the magnitude of \Lya{}  and CMB heating depends on the intensity of the \Lya{} flux, this term is mainly regulated by $f_*$ and $V_c$, while  X-tray heating also  strongly depends on its normalization, $f_X$. To quantify the effect, we calculate  difference (in mK units) in the intensity of the global absorption feature between simulations with the extra heating and reference simulations  without it. 
In Fig. \ref{fig:heating_effect_size} we show the difference versus $f_*$ and $f_X$ for a fixed value of $V_c = 16.5$ and X-ray SED of XRBs  from \citet{fragos13} which peaks at photon energy of $E = 1$ keV (most similar to a power law SED with  $E_{\rm min} = 1$ keV). The results  for a softer power-law SED with $E_{\rm min} =0.1$ keV and $\alpha = 1.5$ and a harder X-ray SED with $E_{\rm min} = 2$ keV and $\alpha = 1.5$  are qualitatively similar.

As expected, the importance of  \Lya{} and CMB heating for the 21-cm signal strongly depends  on the value of $f_X$ and (to a lesser extent) on the shape of the X-ray SED. For high values of $f_X$, X-ray heating is clearly the dominant process, with the vanishing difference  between the full and the reference cases for all the explored values of $f_*$. The difference increases at  lower values of $f_X$ (and develops a dependence on $f_*$), showing that the effects induced by the  \Lya{} background become significant.   For values of $f_X$ below a model-dependent threshold ($f_X\sim 5$ for XRBs), the effect of \Lya{} and CMB heating on the absorption depth is larger than 5 mK. The threshold value is lower/higher for a softer/harder SED as softer X-ray sources are more efficient in heating up  the gas \citep[e.g.,][]{fialkov14a}. We find that  for low $f_X$, \Lya{} and CMB heating can produce an effect of up to $\sim 40$ mK    (for high $f_*$ and $V_c = 16.5$ km s$^{-1}$). 

We also find that for models with weak X-ray heating, the strength of the effect depends on the star formation rate (SFR), which is determined by the minimum mass for star formation (parametrized by $V_c$) and the star formation efficiency, $f_*$. As expected, the effect of \Lya{} and CMB heating is stronger in models with higher $f_*$ and lower $V_c$ as in such cases $x_\alpha$ grows rapidly with redshfit. For models with lower SFR, coupling does not reach saturation by the time reionization eliminates the 21-cm signal. In such cases the absorption depth is relatively small, with or without \Lya{} and CMB heating, and, thus, the absolute difference is small.
Lastly, there is some degree of degeneracy between $f_X$ and $f_*$ in the strength of the effect, owing to the fact that both X-ray heating and the intensity of the \Lya{} background scale as $f_*$ in our model.

\section{The updated landscape of the 21-cm signal}
\label{sec:ranges}

\begin{figure}
    \centering
 \includegraphics[width=0.99\columnwidth]{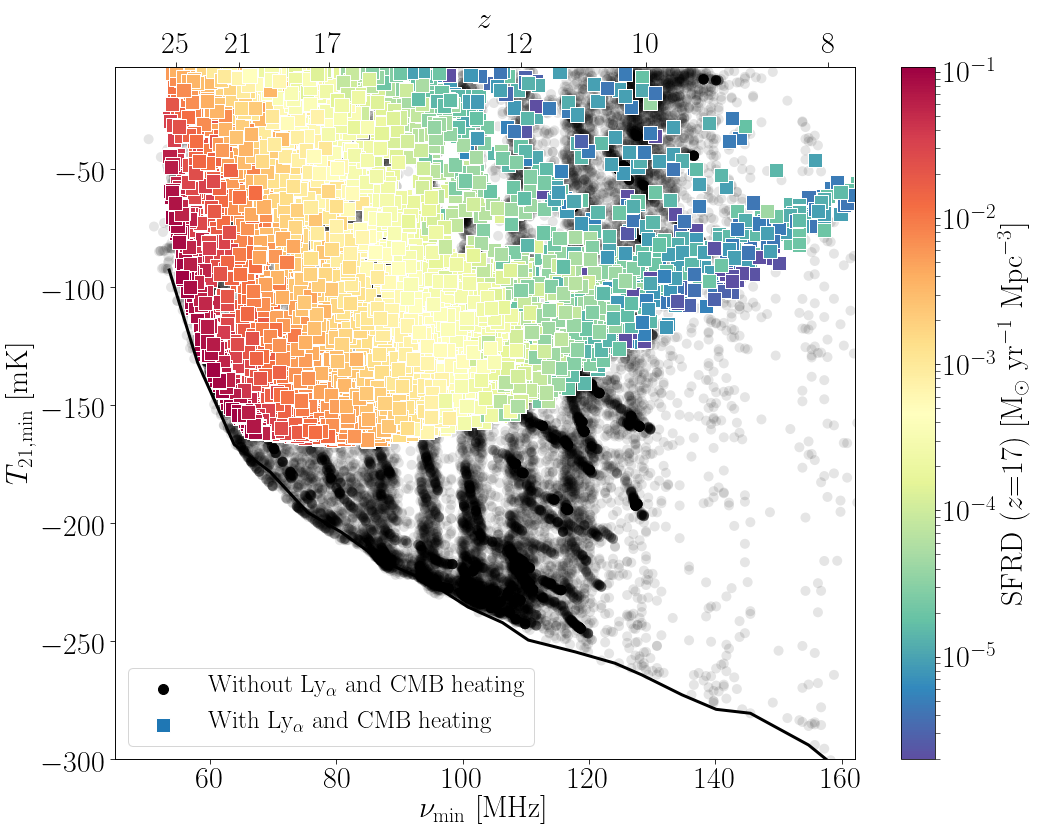}  
    \caption{Absorption depth of the global signal versus the central frequency/redshift (lower/upper $x$-axis) of the absorption trough. Each dot represents a model with a unique combination of the astrophysical parameters \fstar{}, $V_c$, $f_X$, $E_{\rm min}$, $\alpha$, $R_{\rm mfp}$, and $\tau$ (see Table \ref{tab:parameters}). 
    We compare the results obtained with (in color,   for  5968 new astrophysical scenarios) and without  \citep[black, 26886 models, from][]{cohen19}   the newly added effects (\Lya{} and CMB heating, multiple scattering and Poisson fluctuations). 
    The points are colored with respect to the value of  the mean star formation rate density (SFRD) at $z = 17$. }
    \label{fig:tmin_numin}
\end{figure}

\begin{figure*}
    \centering
\includegraphics[width=0.99\columnwidth]{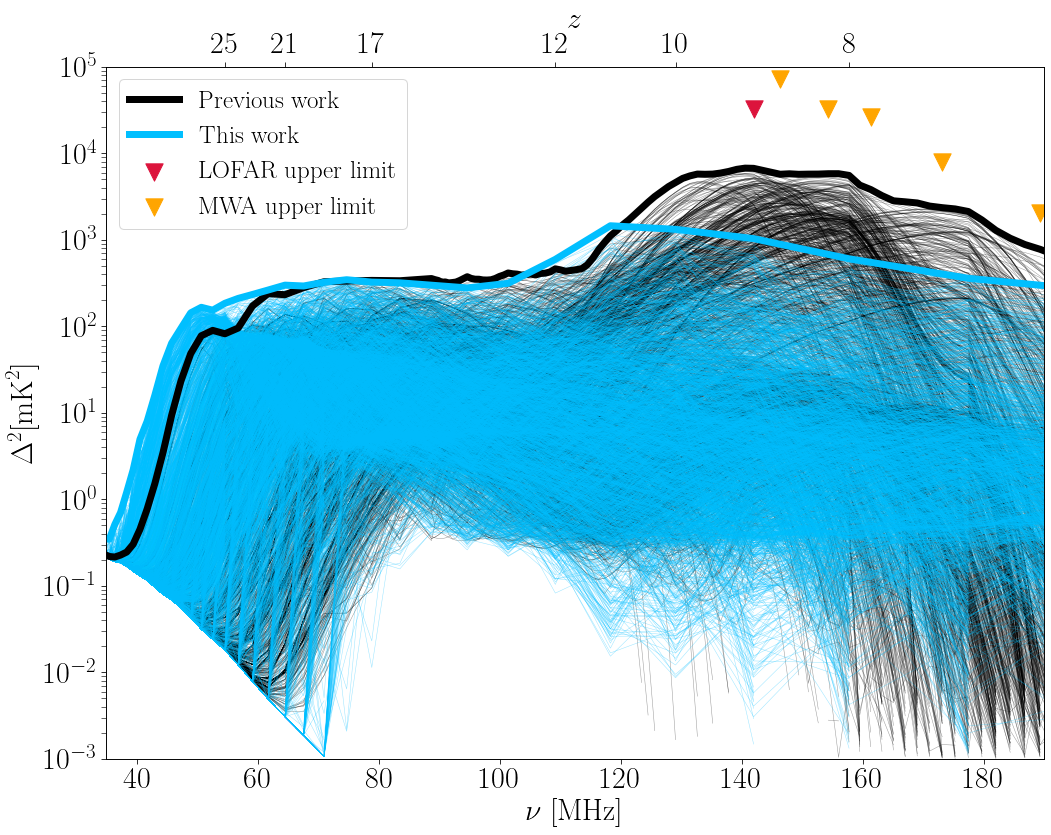} 
\;\; \includegraphics[width=0.99\columnwidth]{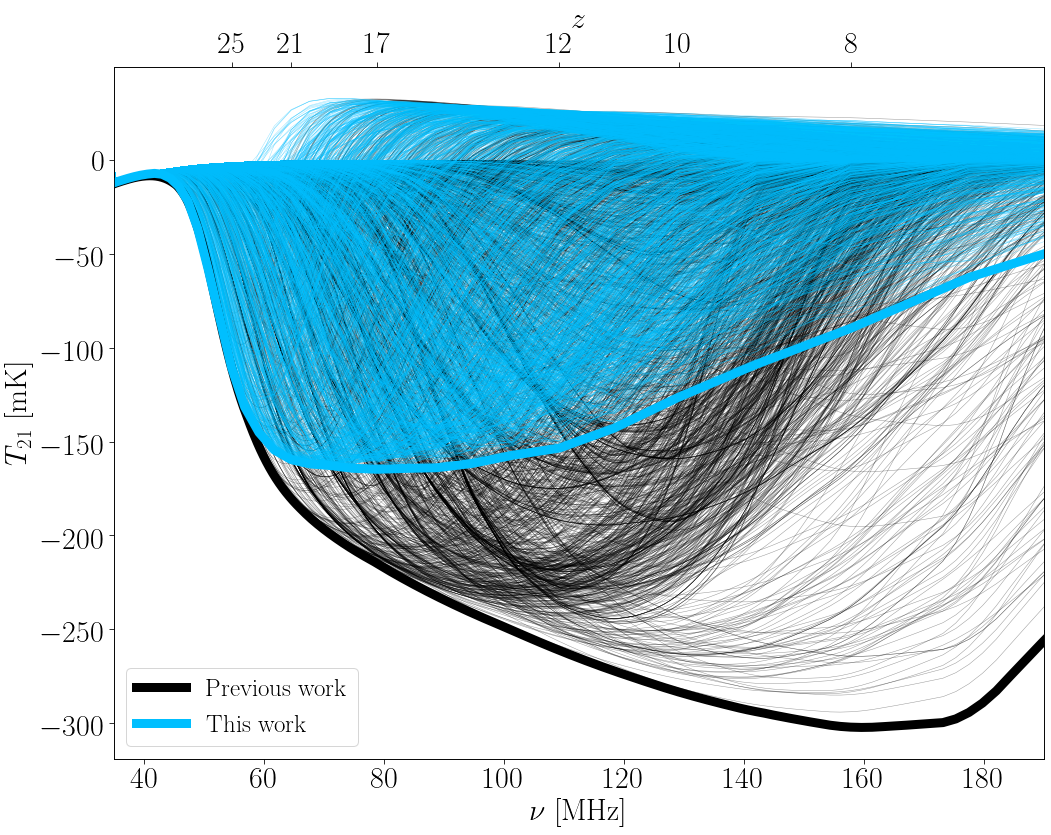} 
    \caption{The updated range of possible 21-cm signals (both the global signal and the power spectrum) in standard astrophysical models. {\bf Left:} Variety of the possible 21-cm power spectra at $k = 0.1$ Mpc$^{-1}$. Blue lines show the results from this work, including the effects of \Lya{} heating, CMB heating, multiple scattering and Poisson fluctuations. Black lines show the results from our previous work \citep{cohen18, cohen19}, where these effects were not included. In each case  2000 randomly selected  models are shown (thin lines) as well as the envelope showing the upper limit at each redshift (thick lines) derived from the entire datasets which cover the parameter ranges listed in Table \ref{tab:parameters} (5968 new models, and 29641/11164 old global signals/power spectra). The red and orange triangle markers show upper limits from LOFAR \citep{LOFAR-EoR:2020} and MWA  \citep{Trott:2020}, respectively. {\bf Right:} Possible 21-cm global signals. Same color coding as for the models on the left. Black lines show the results from our previous work \citep{cohen17, cohen19}. Here the envelopes are lower limits (i.e., indicating the strongest possible absorption at each redshift).}
    \label{fig:ps_env}
\end{figure*}

With the updated framework in place, we explore the entire new landscape of possible 21-cm signals varying the seven model parameters over broad ranges as indicated in Table \ref{tab:parameters}.
In this section we also take into account Poisson fluctuations in the number of halos \citep{reis20a} which were not included in our previous papers. We construct a dataset of 5968 randomly sampled astrophysical scenarios and compare the ranges of both the global signals and the power spectra to the predictions derived using the previous version of our code, where, as discussed above, multiple scattering was taken into account only approximately, while Poisson fluctuations as well as  both \Lya{} and CMB heating were ignored \citep[we use datasets containing  29641 global signals  and 11164 power spectra corresponding to the simulation setup  as in][]{cohen17, cohen18, cohen19}. While both the new and the old datasets are representative, the random sampling might not be optimal and some narrow regions of the parameter space might remain untested. Hence, this  comparison is performed solely for illustration purposes, and  we leave a more thorough exploration of the parameter space (such as needed in order to derive parameter constraints using observational upper limits) for future work. The results are shown in Figs. \ref{fig:tmin_numin} and \ref{fig:ps_env}.

We first consider the range of possible global signals which are strongly affected by the new heating terms. We use the entire dataset of 5968 new full-physics models to explore the distribution of the depth of the absorption feature as a function of the central frequency of the absorption trough (see  Fig. \ref{fig:tmin_numin}) and compare this results to our previous works \citep{cohen17, cohen19}.   We find that, owing to the contribution of the new heating terms, the absorption trough of a global signal in the newly revised models reaches a floor of  $-164.72$ mK at redshifts $z\approx 15-19$.  This is in striking contrast to our previous predictions where the deepest absorption, obtained in the case of the adiabatically cooling Universe without X-ray heating, is a monotonically decreasing function of cosmic time and is $-178.43$ mK  at $z = 19$ and $-216.08$ mK at $z=15$, dropping to even lower values  at lower redshifts (e.g. $-264.40$ mK at $z = 10$). Reduced absorption trough is not the only outcome of the additional heating components. With the extra heating terms, the global 21-cm signals, shown as a function of redshift/frequency on the right panel of Fig.     \ref{fig:ps_env},  are more likely to be narrower, slightly shifted to lower frequencies and exhibit a stronger emission feature compared to the previously considered models.  

The deepest absorption of $-164.72$   mK that we obtain from the new dataset  is $\sim 10\%$ shallower than the signals  demonstrated by \citet{chuzhoy06} and \citet{mittal20}.  The discrepancy could be explained by several factors. First,  these earlier works neglect the effect of CMB heating which adds extra few degrees to the gas temperature leading to a slightly weaker 21-cm signal  (e.g. see Fig. \ref{fig:heating_rates}). Second, we find that  the maximum absorption depth of the 21-cm signal is sensitive to small, $\mathcal{O}(1)$ K, variations in the  gas kinetic temperature. For this reason, the initial conditions used to calculate the gas temperature play an important role. \citet{mittal20} mention that their initial conditions, taken from RECFAST \citep[][]{recfast}, give $T_{\rm K} = 18$ $k$ at $z = 30$. We, on the other hand, use HyRec \citep[][]{hyrec} and assume cosmological parameters from {\it Planck} to create a non-uniform initial temperature cube biased by large scale density fluctuations, which in average gives $T_{\rm K} = 19.86$ K at $z = 30$. 
Finally, we compute the global 21-cm signal as a mean over a large cosmological volume, while in the other works only an averaged calculation was done.

The updated range of possible 21-cm power spectra is shown on the left panel of Fig. \ref{fig:ps_env}. Compared to our old models, here we see a (model and scale dependent) enhancement by a factor of $\sim 2-5$ at high redshifts owing to the effects of multiple scattering and Poisson fluctuations,  and a suppression at low redshift (with peak power lower by a factor of 6.62 at $z=9$ and $k = 0.1$ Mpc$^{-1}$)  due to the additional heating terms.  
While the high-redshift enhancement should make it easier to detect the  signal with the future  SKA,  NenuFAR and HERA at low-frequencies, the reduced range of power spectra at the EoR  redshifts is of a particular importance for the ongoing experiments, such as MWA \citep{Trott:2020},  LOFAR \citep{LOFAR-EoR:2020}, and HERA. Ignoring the effects of the \Lya{} and CMB heating on the predicted low-redshift 21-cm power spectra  will result in underestimating the total energy injection into the  IGM,  overestimating the intensity of the resulting signals, and, consequently, biasing high the estimates of the heating efficiency of X-ray sources when fitting models to data.

\section{Summary}
\label{sec:summary}

In this work we have presented revised predictions for the  21-cm signal that take into account three subtle astrophysical effects induced by the \Lya{} radiative background: \Lya{} heating, CMB heating, and the multiple scattering of \Lya{} photons. The impact of all these effects on the power spectrum of the 21-cm signal is considered here for the first time, while the effects of  \Lya{} heating and CMB heating on the global 21-cm signal have been previously calculated for a handful of high-redshift astrophysical scenarios. In this work we created a large dataset of models, varying the astrophysical parameters in order to cover a broad range of plausible values, and we explored the impact of these  processes on the 21-cm signal. 

We have found that both \Lya{} and CMB heating of the IGM have an important effect once the WF coupling due to \Lya{} photons begins to saturate. CMB heating becomes significant first (when $x_\alpha \sim 5-10$), and \Lya{} heating overtakes it after a delay (when $x_\alpha \sim 50-100$). Both effects are particularly important for astrophysical scenarios with inefficient X-ray heating (typically with $f_X\lesssim 0.1$ for an SED of X-ray binaries). With the formation of the first population of stars and build up of the \Lya{}  background,  these processes heat the cold hydrogen gas, eventually raising its temperature even in the absence of other heating mechanisms. Because these processes inevitably take place as a result of star formation, which also enables the WF coupling that is necessary for a strong 21-cm signal, the popular limiting case of an adiabatically cold IGM with saturated WF coupling is not a realistic bound to consider when comparing models to data. Since the deepest absorption troughs in the global 21-cm spectra and the strongest reionization power spectra are obtained for the coldest possible IGM, \Lya{} and CMB heating significantly change predictions for those signals that are the easiest ones to rule out observationally.  With the newly included effects our models reach an absorption floor at $z\sim15-19$ with a maximum absorption depth of $-165$ mK (considering here standard astrophysical models, without an excess radio background or cooling by millicharged dark matter), in striking contrast with predictions for an adiabatically expanding Universe where the absorption troughs are deeper at lower redshifts \citep[e.g., -264 mK at $z = 10$,][]{cohen16}. Owing to the lower intensity of the 21-cm signals, the 21-cm power spectra are also reduced at low redshifts, with peak power lower by a factor of 6.6 at $z=9$ and $k = 0.1$ Mpc$^{-1}$. This suppression places standard astrophysical models further out of reach of ongoing experiments  \citep[e.g.,][]{Trott:2020, LOFAR-EoR:2020}. 

In contrast with the extra heating terms which suppress the signal, the multiple scattering of \Lya{} photons enhances the power spectrum by a scale-dependent factor of $\sim 2-5$ and renders the cosmic dawn 21-cm signal easier to observe with radio telescopes such as the SKA, NenuFAR and HERA. This effect peaks at the onset of star formation during the coupling transition; however, we find that it can also be significant at lower redshifts in cases with weak X-ray heating.

We conclude that, although created by subtle physical processes, the newly-included effects of the \Lya{} background on the 21-cm signal are important. The landscape of possible 21-cm signals is strongly modified when these effects are included, which should be taken into account for a reliable  estimation of high-redshift astrophysical parameters from observations of both 21-cm power spectra and global signals.

\section*{Acknowledgments}

We acknowledge the usage of the DiRAC HPC. AF was supported by the Royal Society University Research Fellowship. This research was supported for I.R.\ and R.B.\ by the Israel Science Foundation (grant No.\ 2359/20) and by the ISF-NSFC joint research program (grant No.\ 2580/17). We thank Shikhar Mittal and Girish Kulkarni for  useful comments.

\section*{Data availability}
No new data were generated or analysed in support of this research.

\bibliographystyle{mnras}
\bibliography{cosmo}

\appendix

\section{Monte Carlo procedure}
\label{app:mc}
In the Monte Carlo procedure each photon  (emitted at a given frequency $\nu_{\rm em}$ and a given redshift $z_{\rm em}$)  moves in straight lines between scattering events. The distance travelled between two scattering events and the angle of scattering are both random variables. We perform the MC calculation only for photons that are emitted between \Lya{} and \Lyb{}. For higher frequency photons multiple scattering is less important since these photons travel significantly shorter distances in the first place.

 At each step we have the current redshift, frequency, and radial distance from the source: $z_i$, $\nu_i = \frac{1 + z_i}{1 + z_{\rm em}} \nu_{\rm em}$, and $R_i$. To find the redshift of the next scattering event we calculate the optical depth $\tau(z)$ for $z < z_i$ \citep{loeb99}:
 \begin{equation}
 \label{eqn:mc_tau}
     \tau(z) = \frac{\nu_\ast(z_{\rm em})}{\nu_{\alpha}} \left(\frac{\nu_{\rm em}}{\nu_{\alpha}}\right)^{3/2}\left[F\left(\frac{\nu_i}{\nu_{\alpha}} - 1\right) - F\left(\frac{\nu_i}{\nu_{\alpha}}\frac{1+z}{1+z_i} - 1\right)\right],
 \end{equation}
 where
 \begin{multline}
     F(y)= \int dy \frac{(y + 1)^{-5/2}}{y^2} = \\
     \frac{5}{2} \log{\left|\frac{\sqrt{1 + y} + 1}{ \sqrt{1+y} - 1}\right|} - \frac{5y^2 + \frac{20}{3}y + 1}{y(y+1)^{3/2}},
 \end{multline}
 and $\nu_\ast(z) = 5.58 \times 10^{12} \Omega_b h\Omega_m^{-1/2}(1+z)^{3/2} {\rm Hz}$ is the frequency at which the optical depth is equal to one for a photon emitted at $z$. Note that $\tau(z)$ diverges as $\nu_i$ approaches $\nu_{\alpha}$. We  then randomly draw a value for the optical depth at which the photon was absorbed from an $\exp{(-\tau)}$ distribution, and find the redshift at which this value matches $\tau(z)$ from Eqn. \ref{eqn:mc_tau}, this redshift will be $z_{i+1}$.
 
 The comoving distance the photon travelled in this step is then
 \begin{equation}
 \label{eqn:Li}
     L_{i} = \frac{6.0}{\Omega_m^{1/2} h} \left(\sqrt{a(z_{i + 1})} - \sqrt{a(z_{i})}\right) {\rm Gpc}.
 \end{equation}

 Where $a = 1/(1+z)$. To calculate $R_{i+1}$ (the radial distance from the source in the next step) we draw (from a uniform distribution) the direction in which the photon travelled. In comoving coordinates we use Euclidean geometry, so
 \begin{equation}
 \label{eqn:Ri}
     R_{i+1} = \sqrt{R_{i}^2 + L_{i}^2 + 2 R_{i} L_{i} \mu},
 \end{equation}
 where $\mu = \cos{\theta}$, $\theta$ being the angle with respect to the radial direction from the source.
 
 This is repeated until the frequency of the photon is very close to $\nu_{\alpha}$. Here the number of scattering is very large and we use an analytic diffusion approximation from \citet{loeb99} to finish the path of the photon with one final step.  Numerically this happens when $v_{i} < \nu_{\rm min}$, where $\nu_{\min} = \nu_{\alpha} + 0.01 \nu_{\ast}(z_{\rm em})$. We have checked the convergence of our results with respect to the choice of $\nu_{\min}$.
 
 For this step we define the scaled variables 
 \begin{equation}
     \tilde{\nu} = \frac{\nu - \nu_{\alpha}}{\nu_{\ast}(z_{\rm abs})},
 \end{equation}
and
\begin{equation}
    \tilde{L} = \frac{L}{R_{\ast}(z_{\rm abs})}, 
\end{equation}
where
\begin{equation}
    R_{\ast}(z) =6.77 \frac{\Omega_b}{\Omega_m} (1+z) {\rm Mpc}
\end{equation}
is the distance at which $\nu_{\ast}(z)$ is redshifted to $\nu_{\alpha}$. 

According to the approximation from \citet{loeb99}, $\tilde{L}$ could be generated by drawing $\tilde{L}_x, \tilde{L}_y, \tilde{L}_z$ from a Gaussian distribution with mean zero and variance $2 \tilde{\nu} /(9 a_{\rm abs}^2)$, and $\tilde{L} = \sqrt{\tilde{L}_x^2 + \tilde{L}_y^2 + \tilde{L}_z^2}$. We then calculate $L$ for this step, draw a random direction, and obtain the final $R$ with  Eqn. \ref{eqn:Ri}.
Note that since the scattering is elastic, the frequency of the photon changes only due to redshift, so  the emission redshift and frequency defines the absorption redshift $z_{\rm abs}$.

This procedure is repeated for many photons, on grids of emission redshifts $z_{\rm em}$, and frequencies $\nu_{\rm em}$. To get $f_{\rm scatter}(z_{\rm abs}, z_{\rm em}, r)$ (as used in Eqn. \ref{eqn:ja_with_r_dist}) we take all the photons with given $z_{\rm abs}$ and $z_{\rm em}$ and find the fraction of them that travelled a distance $r$. Numerically, for a given  $z_{\rm em}$ (from the MC calculation grid), and  a given $\tilde{z}_{\rm abs}$ used in the 21cm simulation itself, we take all photons that were absorbed between  $\tilde{z}_{\rm abs} + 0.025$ and  $\tilde{z}_{\rm abs} - 0.025$,  and bin them on a grid of distances. We have checked the convergence of our results with respect to the redshift bins.  For each bin we interpolate the fraction of photons  between the  different $z_{\rm em}$ to the simulation value $\tilde{z}_{\rm em}$.

 We fit the following (purely phenomenological) function  to the distance distribution $n(r)$ in logarithmic bins:
\begin{multline}
 \label{eqn:fit_func}
    \log\left( \frac{d n}{d log(r)}\right) = \gamma  \log\left( \frac{r}{r_{\rm min}}\right) \times \left(\frac{1}{1 +  \left( \frac{r}{r_{\rm mid}}\right)^{-\beta/2}  }\right)^2 + \\ A \times \left( \frac{r}{ r_{\rm mid} e^{-\beta}  }  \right) ^{-2\alpha^2} 
 \end{multline}
 for $r_{\rm min} < r < r_{\rm SL}$, and 
 \begin{equation}
      \log\left( \frac{d n}{d log(r)}\right)= 1
 \end{equation}
 otherwise.  $r_{\rm SL}$ is the straight line distance, and $r_{\rm min}, r_{\rm mid}$, $\gamma$, $\beta$, $\alpha$, and $A$ are free parameters.

 The first term represents a power law  $dn/dr \sim r^{(\gamma -1)}$, where the $-1$ comes from the logarithmic bins. Note that the number of photons in a simulation cell goes like $r^{(\gamma -3)}$ as the photons are distributed over spherical shells with $V\sim r^{2}$. The $r_{\rm min}$ parameter is used only in the fitting procedure and does not represent an actual minimum distance (it represents the minimum distance that will receive a photon given the number of iterations of the MC code). The power law for $dn/dr$ is  thus extrapolated to smaller distances. The second term is a cutoff at distance $r_{\rm mid}$ with log-space width $\sim \beta$. The last term is  accounting for accumulation of photons close to the maximum distance (which is located roughly at the start of the drop, i.e, at $r_{\rm mid} \times e^{- \beta}$). The choice of function is  phenomenological and was found to provide a good fit to the distance distribution over the entire range of emission and absorption redshifts ($z =6-40$ in our simulation).
 
 $f_{\rm scatter}(z_{\rm abs}, z_{\rm em}, r)$, used in Eqn. \ref{eqn:ja_with_r_dist}, is equal to  $dn/dr$ extrapolated to $r < r_{min}$ and normalized  so that $\int dr f_{\rm scatter}(z_{\rm abs}, z_{\rm em}, r) = 1$. We have shown example results of this procedure in \citet{reis20a}.

Computationally, $f_{\rm scatter}(z_{\rm abs}, z_{\rm em}, r)$ serves to calculate the new window functions in real space
\begin{equation}
    W(z_{\rm abs}, z_{\rm em}, {\textbf r}) = \frac{f_{\rm scatter}(z_{\rm abs}, z_{\rm em}, r)}{4 \pi r^2},
\end{equation}
which we use in the simulation instead of the spherical shell window functions\footnote{For comparison, for  spherical shell window functions,  $W(z_{\rm abs}, z_{\rm em})$ is non zero only at a single value of $r$ given by Eqn. \ref{eq:straight_line_distance}.}.
Note that these window functions are normalized, $\int d\textbf{r} W = 1$. The second integral in Eqn. \ref{eqn:ja_with_r_dist}  can then be calculated by convolving the window function with the emissivity:
\begin{multline}
       \int dr f_{\rm scatter}(z_{\rm abs}, z_{\rm em}, r) \epsilon( \nu'_n(z_{\rm abs}, z_{\rm em}), z_{\rm em} )_{r, \textbf{x}} = \\ 
       \int d\textbf{r} \frac{f_{\rm scatter}(z_{\rm abs}, z_{\rm em}, r)}{4 \pi r^2} \epsilon( \nu'_n(z_{\rm abs}, z_{\rm em}), z_{\rm em}, \textbf{x} + \textbf{r}) =
       \\ 
       \int d\textbf{r} W(z_{\rm abs}, z_{\rm em}, \textbf{r})  \epsilon( \nu'_n(z_{\rm abs}, z_{\rm em}), z_{\rm em}, \textbf{x} + \textbf{r}) =
       \\ W(z_{\rm abs}, z_{\rm em}) \circledast \epsilon( \nu'_n(z_{\rm abs}, z_{\rm em}), z_{\rm em}).
\end{multline}

\section{Inhomogeneity of \Lya{} heating and its effect on the fluctuations of the 21-cm signal}
\label{app:PS}

\begin{figure}
    \centering
 \includegraphics[width=0.99\columnwidth]{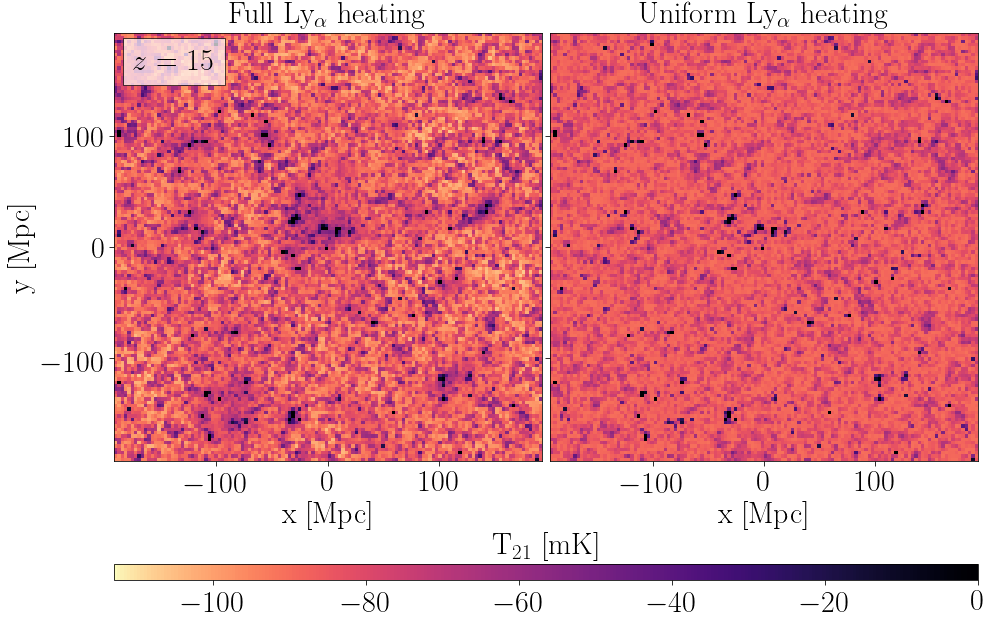} \\ [6 pt]  
 \includegraphics[width=0.99\columnwidth]{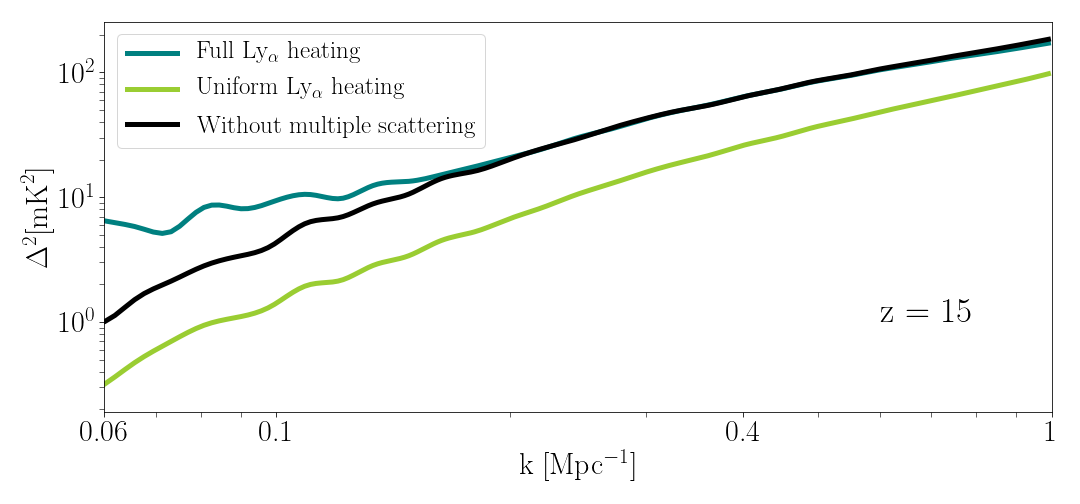}
    \caption{The effect of \Lya{} heating  on the 21-cm fluctuations for an example model with $V_c = 16.5$ km s$^{-1}$, $f_* = 0.251$ and $f_X = 0.001$ (the blue curve in the top panel of Fig. \ref{fig:power_same_vc_diff_fs}).
    {\bf Top:} a slice of the 21-cm brightness temperature at $z=13$ for the full case ({\bf left}) and the case with an artificially uniform \Lya{} heating rate ({\bf right}). The color bar shows the values of the brightness temperature in mK. Note that the regions with zero brightness temperature are ionized (these are the highest density regions).   {\bf Bottom:  } The corresponding power spectra at $z=15$ for the full case (dark green) and the case with a uniform \Lya{} heating rate (light green). We also show a case without the multiple scattering of \Lya{} photons (dark blue). }  
    \label{fig:lya_heating_slice}
\end{figure}

The  \Lya{} contribution to heating is expected to be smooth, as for this component to have a significant effect on the 21-cm signal, strong  \Lya{} background created by multiple sources is required. Here we test this hypothesis by isolating the effect of \Lya{} heating and  comparing our full simulation to a test case where  \Lya{} heating is set to be spatially homogeneous. The test case is constructed so that the \Lya{}  contribution to the heating rate, $ \epsilon_{\rm Ly_{\alpha}}$ (Eqn. \ref{eqn:heating_rate}), is spatially uniform with the same mean value as in the original simulation. 
Examples of 21-cm images with the smooth and  the non-uniform \Lya{}   heating rates are shown in Fig. \ref{fig:lya_heating_slice}.  In both cases the simulation parameters are  $V_c$ = 16.5 km s$^{-1}$, \fstar{} = 0.2507, $f_X$ = 0.001 (so that \Lya{} heating is the dominant heating mechanism), and $\zeta = 30$. The images are shown at $z=15$, which is the epoch at which heating fluctuations dominate.

As can be seen from the figure,  \Lya{} heating  is intrinsically non-uniform and enhances fluctuations in the 21-cm signal.
In the figure the regions with low brightness temperature are relatively high density regions (the highest density regions are already ionized).  Due to the fluctuations in the \Lya{} heating rate, these regions have lower kinetic temperature and thus a lower brightness temperature (compared to the test case with a uniform \Lya{} heating rate). To understand this result  it is important to realize that \Lya{} heating is stronger in {\it low} density regions. This can be non-intuitive, since radiation sources are localized in high density regions contributing to a larger number of  \Lya{} photons in these regions. However, \Lya{} heating depends on the ratio between the number of \Lya{} photons and the number of baryons. This dependence on the number of baryons reduces the effect of \Lya{} heating in high density regions compared to the low density regions. In addition, as discussed above, \Lya{} heating is determined by the competition between continuum photons which heat the gas and injected photons which cool the gas. The fact that continuum photons travel larger distances relative to injected photons\footnote{This effect is  included in the pre-calculated coefficients (the same place where the spectrum of the sources comes into play). For each absorption redshift we calculate the maximum emission redshift  (or horizon)  for each Lyman resonance. For \Lya{} the horizon is the largest, and becomes smaller for \Lyb{}, \Lyg{} etc.   } also reduces the heating rate near the sources (that is, in high density regions) compared to far away from the sources (in low density regions).

Finally, the effect of \Lya{} heating fluctuations can also be seen in the 21-cm power spectrum (bottom panel of Fig. \ref{fig:lya_heating_slice}), which is enhanced on all scales probed by our simulation by a factor of $\sim 2$ at $z=13$ for this specific choice of astrophysical parameters. Fig. \ref{fig:lya_heating_slice} also show a case without multiple scattering of \Lya{} photons. This examples illustrates that when \Lya{} heating is the dominant heating mechanism, multiple scattering affects 21-cm fluctuations during the heating transition as well, and not only in the high-redshift regime.

\section{CMB and \Lya{} heating examples}
\label{app:heating_rates}

\begin{figure*}
    \centering
  \includegraphics[width=0.99\textwidth]{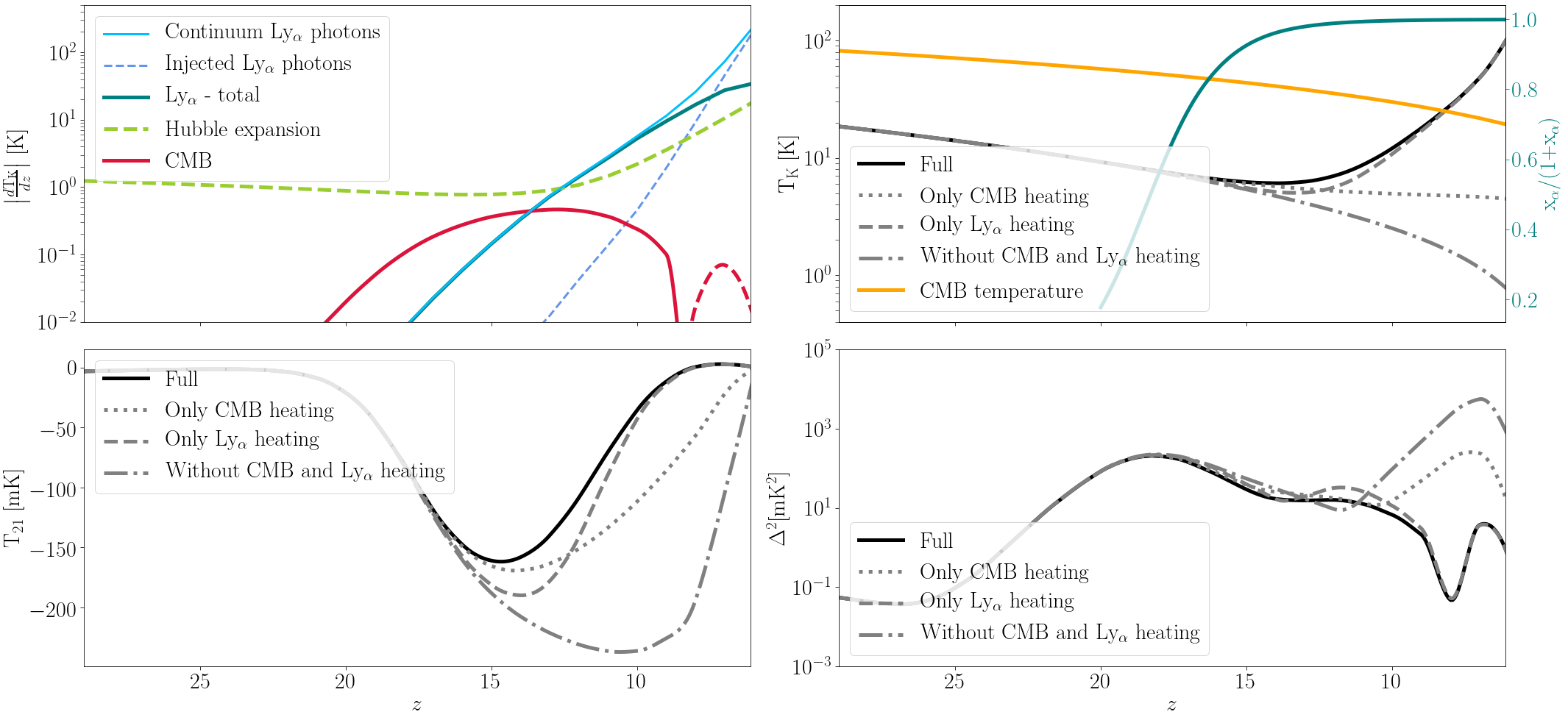}  \\ [6pt]
    \caption{
    Same as Fig. \ref{fig:heating_rates} but for an astrophysical scenario with  $f_X = $ 0 (no X-rays) , $f_*$ = 0.2, and $V_c$ = 50 km s$^{-1}$. } 

    \label{fig:heating_rates_extra_1}
\end{figure*}

\begin{figure*}
    \centering
  \includegraphics[width=0.99\textwidth]{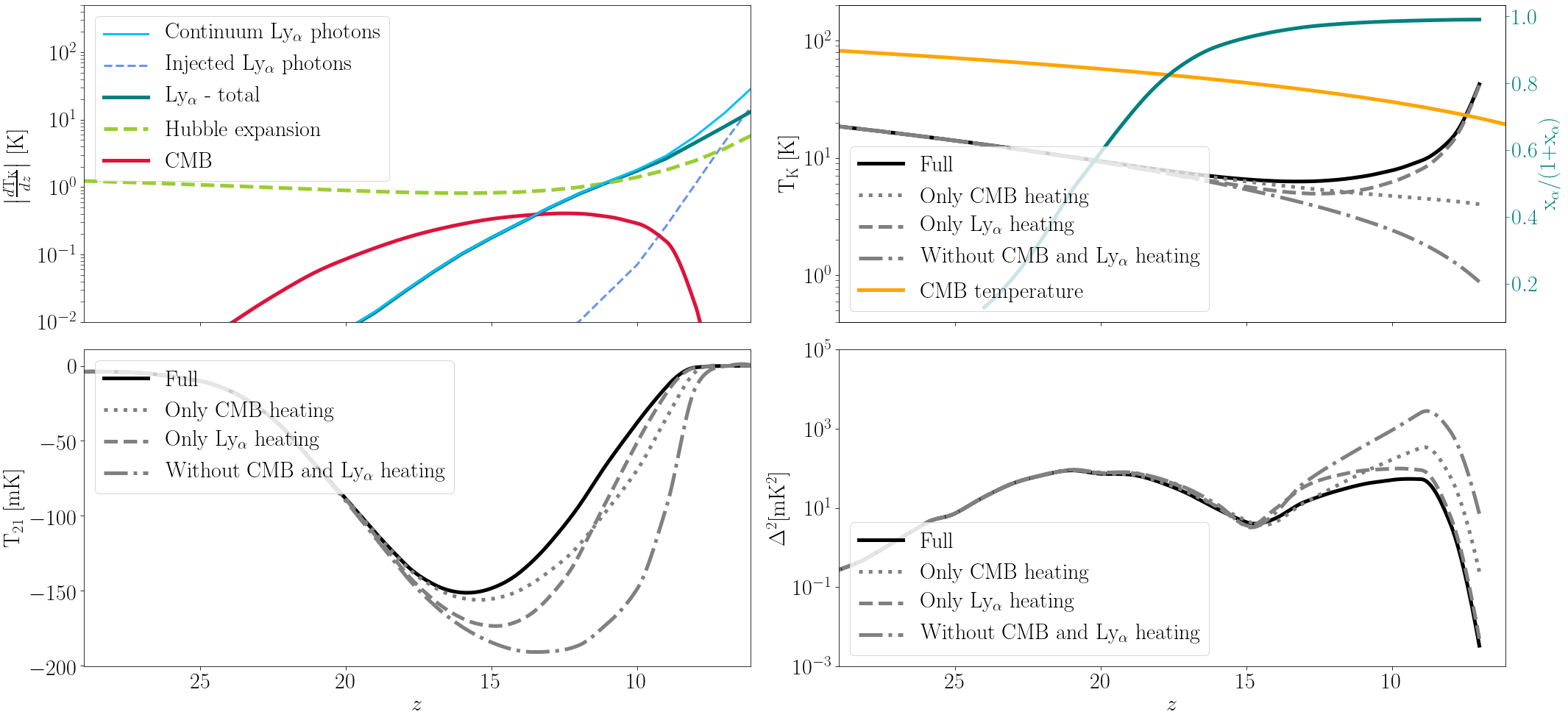}  \\ [6pt]
    \caption{
    Same as Fig. \ref{fig:heating_rates} but for an astrophysical scenario with  $f_X = $ 0 (no X-rays) , $f_*$ = 0.01, and $V_c$ = 16.5 km s$^{-1}$. } 

    \label{fig:heating_rates_extra_2}
\end{figure*}

In Fig. \ref{fig:heating_rates_extra_1} and Fig. \ref{fig:heating_rates_extra_2} we show the same quantities as in Fig. \ref{fig:heating_rates} for two additional astrophysical models. In Fig. \ref{fig:heating_rates_extra_1} we show a case with high circular velocity (corresponding to a high minimum mass for star formation), and in Fig. \ref{fig:heating_rates_extra_2} we show a case with low star formation efficiency. In both cases, as in the case shown in the main text, we see a significant suppression of the global signal and the EoR power spectrum due to \Lya{} and CMB heating.

\bsp	
\label{lastpage}
\end{document}